\documentclass[useAMS,usenatbib]{mn2e}
\usepackage{epsfig}
\usepackage{amsmath}

\newcommand{\be}{\begin{equation}}
\newcommand{\beq}{\begin{equation}}
\newcommand{\ba}{\begin{eqnarray}}
\newcommand{\ee}{\end{equation}}
\newcommand{\eeq}{\end{equation}}
\newcommand{\ea}{\end{eqnarray}}

\newcommand{\hs}{\hspace{1mm}}

\newcommand\ion[2]{#1$\;${\scshape{#2}}}%         

\def\lsim{~\rlap{$<$}{\lower 1.0ex\hbox{$\sim$}}}

\def\gsim{~\rlap{$>$}{\lower 1.0ex\hbox{$\sim$}}}

\title[The clustering of Ly$\alpha$ emitters]{Non-Gravitational Contributions to the Clustering of Ly$\alpha$ Selected Galaxies: Implications for Cosmological Surveys}

\author[Wyithe \& Dijkstra]{J. Stuart B. Wyithe$^1$ and Mark Dijkstra$^2$\\$^1$
School of Physics, University of Melbourne, Parkville, Victoria 3010,
Australia\\$^2$ Max Planck Institute fur Astrophysik, Karl-Schwarzschild-Str. 1, 85741 Garching, Germany
\\Email: swyithe@physics.unimelb.edu.au}

\begin{document}

%\date{\today}
%\pagerange{\pageref{firstpage}--\pageref{lastpage}} \pubyear{2006}

\maketitle

\label{firstpage}
\begin{abstract}

We show that the dependence of Ly$\alpha$ absorption on environment leads to significant non-gravitational features in the redshift space power-spectrum of Ly$\alpha$ selected galaxies. We derive a physically motivated fitting formula that can be included in clustering analyses, and use this to discuss the predicted features in the Ly$\alpha$ galaxy power-spectrum based on detailed models in which Ly$\alpha$ absorption is influenced by gas infall and/or by strong galactic outflows. We show that power-spectrum measurements could be used to study the astrophysics of the galaxy-IGM connection, and to measure the properties of outflows from star-forming galaxies.  
Applying the modified redshift space power-spectrum to a Ly$\alpha$ survey with parameters corresponding to the planned Hobby-Eberly Telescope Dark Energy Experiment (HETDEX), we find that 
the dependence of observed Ly$\alpha$ flux on velocity gradient and ionising background may compromise the ability of Ly$\alpha$ selected galaxy redshift surveys to constrain cosmology using information from the full power-spectrum. This is because the effects of fluctuating ionizing background and velocity gradients effect the shape of the observed power-spectrum in ways that are similar to the shape of the primordial power-spectrum and redshift space distortions respectively. We use the Alcock-Paczynski test to show that without prior knowledge of the details of Ly$\alpha$ absorption in the IGM, the precision of line-of-sight and transverse distance measurements for HETDEX will be $\sim1.3-1.7\%$, decreased by a factor of $\sim1.5-2$ relative to the best case precision of $\sim0.8\%$ available in a traditional galaxy redshift survey. 
We specify the precision with which modelling of Ly$\alpha$ radiative transfer must be understood in order for HETDEX to achieve distance measurements that are better than 1\%.

\end{abstract}


\section{Introduction}

The Ly$\alpha$ emission line of galaxies provides a primary observable for discovering high redshift galaxies \citep[e.g.][]{kashikawa2006,ouchi2010,Iye06,Lehnert10}, for studying their starformation and inter-stellar medium \cite[e.g.][]{verhamme2006,Dessauges10,Steidel11}, and for studying the ionisation state of the inter-galactic medium or IGM \citep[e.g.][]{Haiman99,Malhotra04,kashikawa2006,dijkstra2007}. In addition to measuring the luminosity function  of Ly$\alpha$ emitting galaxies \citep[e.g.][]{shimasaku2006,kashikawa2006,Ouchi08,Cassata11,ouchi2010,Blanc11}, samples have recently become large enough to enable studies of Ly$\alpha$ galaxy clustering \citep[e.g.][]{gawiser2007,kovac2007,orsi2008,guaita2010,ouchi2010}. These clustering studies yield complementary information to the luminosity function, since clustering of galaxies can provide a direct estimate of the halo mass, independent of the life-time of star-burst activity. Comparison with the luminosity function therefore provides an avenue to estimate the overall efficiency and duration of star-formation activity \citep[e.g.][]{nagamine2010}. 

Unlike populations of Ly-break galaxies that are selected via broad band photometry, the observed brightness of a Ly$\alpha$ emitter is sensitive to its local extra-galactic environment. In particular, Ly$\alpha$ radiation that escapes from galaxies can be scattered out of the line-of-sight by neutral hydrogen atoms in inter-galactic medium (IGM) surrounding the galaxy. For example, prior to the completion of reionisation the strength of the damping wing in the Ly$\alpha$ absorption line means that Ly$\alpha$ galaxies should be more easily detected inside the HII regions that are thought to have been generated by clustered early star-forming galaxies \citep[e.g.][]{santos2004,wyithe2005,Mesinger08b}, leading to enhanced clustering that could provide a signature of patchy reionisation \cite[][]{mcquinn2007b,iliev2008,Mesinger08}. 

Following the conclusion of reionisation, the fraction of radiation that is scattered out of the line-of-sight is dependent on resonant absorption within the highly ionised IGM. In this regime, the fraction of Ly$\alpha$ flux that is transmitted to the observer is dependent on quantities like the infall velocity,  the local overdensity of mass,  and the ionising background. Any environmental dependence of observed flux at fixed intrinsic luminosity therefore leads to an environmental dependence of the host halo mass of observed galaxies at fixed observed flux. This in turn leads to a dependence of the observed galaxy density on environment that differs to that expected from galaxy bias. Since clustering studies are performed in flux limited surveys, the net result is a modification of the observed clustering of Ly$\alpha$ selected galaxies \citep[][]{zheng2010}. This modification of the observed clustering is non-gravitational.     

Galaxies are thought to provide a (biased) tracer of the density field, and hence their clustering can be used to infer the statistical properties of the mass-density field on the large scales that probe cosmology \citep[e.g.][]{cole2005,eisenstein2005,percival2007,okumura2008,gaztanaga2008,reid2010}. The gravitational contributions to this clustering can be studied analytically, providing qualitative interpretation of the observations \cite[e.g.][]{sheth2001}. However the precision of modern galaxy redshift surveys has meant that N-body analyses are required to understand the observations in detail \citep[e.g.][]{tinker2006,eisenstein2007,seo2008}. While gravitational effects are thought to be the only mechanism influencing clustering at an observable level in traditional galaxy redshift surveys, studies of clustering among Ly$\alpha$ selected galaxies will need to also include non-gravitational contributions. 
For example, the effect of peculiar velocity gradients on the observed clustering of Ly$\alpha$ selected galaxies will not be the same as in a usual galaxy redshift survey  \citep[][]{zheng2010}. This is because both the observed luminosity and the observed redshift space density of the Ly$\alpha$ galaxies depend on velocity gradient.  

\citet[][]{zheng2010}  employed a numerical simulation and calculated the full radiative transfer of Ly$\alpha$ photons to study the clustering of Ly$\alpha$ selected galaxies at $z=5.7$. They argue that the effect of velocity gradients leads to line-of-sight clustering that is suppressed, in contrast to the enhancement seen in traditional galaxy redshift surveys \citep[e.g.][]{kaiser1987,peacock2001}. In addition to identifying this new clustering effect,  \citet[][]{zheng2010} argue that clustering will be enhanced transverse to the line-of-sight, leading to a clustering amplitude that is increased relative to expectations in the absence of Ly$\alpha$ transmission effects.

An important difference between the study of \citet[][]{zheng2010} and previous related work \citep[e.g.][]{santos2004,DijkstraIGM,orsi2008,dayal2009} is the implementation of full radiative transfer within their simulation \citep[][]{zheng2010a}. Many previous studies have utilised a model in which the fraction of Ly$\alpha$ photons that reach the observer is equal to $\exp(-\tau)$, where $\tau$ is the integrated Ly$\alpha$ optical depth along the line of sight. However \citet[][]{zheng2010}  argue that photons are scattered back into the line-of-sight along directions of low density, and so must be included in addition to the directly transmitted photons accounted for in the $\exp(-\tau)$ model. They argue that while the $\exp(-\tau)$ model provides a qualitative explanation of observed Ly$\alpha$ galaxy properties, quantitative differences are found which require full radiative transfer to interpret. 

The issue is partially related to the size, and hence the surface brightness, of the region from which Ly$\alpha$ photons are scattered to the observer. This size is very sensitive to the properties of the gas within the virial radius of a galaxy, and also depends on the intrinsic broadness of the line \citep[][]{laursen2011}.  In a recent paper \citet[][]{laursen2011} have modeled sight-lines to galaxies at lower redshift and at much higher resolution than are available in the simulations of \citet[][]{zheng2010a}. In contrast to \citet[][]{zheng2010}, \citet[][]{laursen2011}  find that the $\exp(-\tau)$ model provides an excellent description of the fraction of transmitted flux that they compute from their high resolution radiative transfer simulations. However \citet[][]{laursen2011} are not able to study the effect of transmission on Ly$\alpha$ galaxy clustering.    

We predominantly consider clustering of Ly$\alpha$ galaxies at $z\la3$. Following \citet[][]{laursen2011}, we therefore employ an  $\exp(-\tau)$ model for Ly$\alpha$ transmission, and use this to explore the possible effect of environmental dependence of Ly$\alpha$ transmission on the clustering of Ly$\alpha$ selected galaxies. This enables us to draw on the extensive prior work evaluating the possible effects of different astrophysical effects including infall and star-formation rates \citep[][]{DijkstraIGM}, and galactic winds and outflows \citep[][]{Ahn03,verhamme2008,dijkstra2010}. The detailed analytic models we use in this work are in good agreement with the simulations of \citet[][]{laursen2011}. Since we do not know the phenomena that are most important in setting the observed flux of Ly$\alpha$ emitting galaxies, the goal of this work is not to produce a detailed model for the Ly$\alpha$ luminosity function itself, or to predict the clustering amplitude \citep[e.g.][]{orsi2008,Shimizu11}. Rather, we aim to understand the effect that fluctuations in transmission will have on the shape of the power-spectrum as a function of scale and direction, as well as on the clustering amplitude. 

Most studies of clustering among Ly$\alpha$ emitting galaxies at high redshift have concentrated on applications related to probing the history of reionisation and the sources responsible for that event. However recent attention has also focussed on obtaining large samples of Ly$\alpha$ emitters to use as tracers of the density field \citep[][]{hill2004b,hill2008a}, with application to cosmological distance measure and probes of dark energy at redshifts not previously studied (Hobby-Eberly Telescope Dark Energy Experiment, HETDEX). In this paper we discuss the implications of non-gravitational contributions to the observed clustering of Ly$\alpha$ emitting galaxies for studies of the shape, amplitude and angular dependence of the Ly$\alpha$ galaxy power-spectrum. To this end we provide a simple framework that allows the influence of non-gravitational effects from Ly$\alpha$ transmission to be evaluated with respect to the available precision of cosmological constraints. As part of our study we determine the detail with which the astrophysics governing the observed properties and flux of Ly$\alpha$ emitters must be understood, in order for a large survey like HETDEX to achieve its theoretical performance in measurement of cosmological parameters.

Our paper is set out as follows. In \S~\ref{model} we construct a linear theory model for the power-spectrum of Ly$\alpha$ selected galaxies, which is generalised to be applicable to any particular model of Ly$\alpha$ transmission. We next discuss a simple analytic model of transmission (\S~\ref{analmodel}), in which a fraction $F$ of the Ly$\alpha$ line is subject to optical depth $\tau_0$. We provide an analytic formula for the resulting redshift space and spherically averaged power-spectra, which we use to discuss the physical origin of different features in the power-spectrum. In \S~\ref{detailedmodel} we investigate the effect of transmission fluctuations on clustering using detailed, previously published models of Ly$\alpha$ transmission. Having determined the likely extent of non-gravitational contributions to the observed clustering, we present an analysis of the likely impact of possible fluctuations in Ly$\alpha$ transmission on the measurements of cosmological parameters in the HETDEX survey (\S~\ref{HETDEX}). As a specific example we compute the Alcock-Paczynski effect~\citep[][]{Alcock1979}. In \S~\ref{LYAconstraints}, we turn this analysis around and discuss the precision that will be available for constraints on Ly$\alpha$ transmission models. We present our conclusions in \S~\ref{conclusions}. In our numerical examples, we adopt the standard set of cosmological parameters \citep[][]{komatsu2011}, with values of $\Omega_{\rm m}=0.24$, $\Omega_{\rm b}=0.04$ and $\Omega_Q=0.76$ for the matter, baryon, and dark energy fractional density respectively, and $h=0.73$, for the dimensionless Hubble constant.

\section{Model for the Clustering of Ly$\alpha$ Selected Galaxies} 
\label{model}

We begin by briefly summarising the basic theory for absorption of Ly$\alpha$ photons from galaxies in the IGM, and then present a simple derivation of galaxy bias in linear theory. We use these as a basis to discuss the effects of fluctuations in the transmission of Ly$\alpha$ flux through the IGM on the observed clustering of Ly$\alpha$ emitters. Our derivation of fluctuations is similar to the analytic model presented in \citet[][]{zheng2010}.

\subsection{Ly$\alpha$ absorption in the IGM}

The transmission of Ly$\alpha$ photons \citep[e.g.][]{DijkstraIGM} is summarised briefly here to provide context for the new clustering calculations. The total opacity seen by a photon initially at frequency $\nu$ is 
\begin{equation}
\label{taunu}
\tau(\nu)=\int_{r_{\rm vir}}^\infty dr\,n_{\rm H}(r)x_{\rm H}(r) \sigma_{\rm Ly}\left(\nu\times\left[1+{v_z(r)}/{c}\right]\right),
\end{equation}
where the Ly$\alpha$ absorption cross-section is written as $\sigma_{\rm Ly}(\nu)$, and the gas at distance $r$ greater than the virial radius $r_{\rm vir}$ has line-of-sight velocity $v_z(r)$. In this expression $x_{\rm H}$ is the fraction of hydrogen in atomic form, given at photoionisation equilibrium in the optically thin limit by 
\begin{equation}
x_{\rm H} = \frac{n_{\rm H}\alpha_{\rm rec}}{\Gamma},
\end{equation}
where $n_{\rm H}$ is the number density of hydrogen nuclii, $\Gamma$ is the photoionisation rate and $\alpha_{\rm rec}$ is the Case-B recombination coefficient, $\alpha_{\rm rec} = 4.2\times10^{-13} (T_{\rm gas}/10^4\mbox{K})^{-0.7}\mbox{cm}^3\mbox{s}^{-1}$ \citep[e.g.][]{hui1997}. Substituting we find
\begin{equation}
\label{tauint1}
\tau(\nu)= \int_{\rm r_{\rm vir}}^\infty dr\, \frac{n^2_{\rm H}(r)\alpha_{\rm rec}}{\Gamma(r)} \sigma_{\rm Ly}\left(\nu\times\left[1+{v_z(r)}/{c}\right]\right).
\end{equation}
The total transmission is found by integrating over the flux density $J(\nu)$ in the Ly$\alpha$ line
\begin{equation}
\label{tranint}
\mathcal{T} = \frac{\int_{-\infty}^{\infty} d\nu J(\nu) e^{-\tau(\nu)}}{\int_{-\infty}^{\infty} d\nu J(\nu)}.
\end{equation}

Approximating the Ly$\alpha$ scattering cross-section as a delta function $\sigma_{\rm Ly}(\nu^\prime)\sim\sigma_{\rm Ly,tot}\delta(\nu^\prime-\nu_{\rm Ly})$, where $\sigma_{\rm Ly,tot}\equiv \int \sigma_{\rm Ly}d\nu=f_{\alpha}\frac{\pi e^2}{m_e c}$  \citep{RL79}\footnote{Here, $f_{\alpha}=0.4167$ denotes the oscillator strength for the Ly$\alpha$ transition, and $e$ and $m_e$ denote the charge and mass of the electron, respectively.} and  $\nu^\prime(r) = \nu\times\left[1+{v_z(r)}/{c}\right]$, we can re-write equation~(\ref{tauint1}) as 
\begin{eqnarray}
\label{tauint}
\tau(\nu)&\approx& \int_{\rm r_{\rm vir}}^\infty dr\, \frac{n^2_{\rm H}(r)\alpha_{\rm rec}}{\Gamma(r)} \frac{\sigma_{\rm Ly,tot}\delta(r-r_{\rm Ly})}{\frac{1}{c}\frac{dv_z}{dr}} \\ \nonumber &=&\frac{cn^2_{\rm H}(r_{\rm Ly})\alpha_{\rm rec}\sigma_{\rm Ly,tot}}{\Gamma(r_{\rm Ly})\left.\frac{dv_z}{dr}\right|_{r_{\rm Ly}}} ,
\end{eqnarray}
where $\nu_{\rm Ly}=\nu[1+{v_z(r_{\rm Ly})}/{c}]$, in which $r_{\rm Ly}$ denotes the radius at which the photon is at resonance in the frame of the gas. Equation~(\ref{tauint}) illustrates the important point that the probability of Ly$\alpha$ scattering is proportional to the total number of hydrogens that the photon encounters per unit velocity along the line-of-sight. The optical depth is therefore proportional to the inverse of the line-of-sight velocity gradient $(\frac{dv_z}{dr})^{-1}$, in addition to the more obvious dependencies of density squared and inverse of ionisation rate. Thus we find
\begin{equation}
\label{taurelation}
\tau \propto \frac{\rho^2}{\Gamma T^{0.7}\frac{dv_z}{dr}} \propto \frac{\rho^{2-0.7(\gamma-1)}}{\Gamma\frac{dv_z}{dr}},
\end{equation}
where in the second proportionality we have assumed a power-law relation between density and temperature of $T\propto \rho^{\gamma-1}$, with  the polytropic index $\gamma=1.4$ \citep[][]{hui1997}. 

\subsection{Galaxy bias and fluctuations in the number density of galaxies}
The likelihood of observing a galaxy at a random location is proportional to the local number density of galaxies. The likelihood of observing a galaxy within a region of large-scale overdensity is therefore equal to the ratio of the number density of halos $n(\delta)$ in a region of large-scale over-density $\delta$ to the number density of halos in the background universe ($\bar{n}$). This ratio has been used to derive galaxy bias for small values of $\delta$ \citep[][]{mo1996,sheth2001}. For example, in the \citet[][]{press1974} formalism we write
\begin{eqnarray}
\label{analbias}
\nonumber 
\frac{dn(\delta)/dM}{d\bar{n}/dM} &=&
(1+\delta)\left[\frac{d\bar{n}}{dM} +
\left.\frac{d^2n}{dMd\nu}\right|_{\nu}\frac{d\nu}{d\delta}\delta\right]\left[\frac{d\bar{n}}{dM}\right]^{-1}
\\&\sim&
1+\delta\left(1+\frac{\nu^2-1}{\sigma(M)\nu}\right)\equiv1+\delta b,
\end{eqnarray}
where $(dn/dM)(\bar{\nu})$ and $(dn/dM)(\nu)$ are the average and perturbed mass functions, $\nu \equiv (\delta-1.69)/\sigma(M)$,  $\sigma(M)$ is the variance in the density field smoothed on a mass-scale $M$ at redshift $z$, and $b$ is the bias factor. 
 At small values of large scale overdensity $\delta$, and in the absence of effects that influence the relation between observed flux and halo mass, the number density of galaxies is proportional to $[1+\delta b(M,z)]$. 

\subsection{The Ly$\alpha$ emitter power spectrum}

In this section we estimate the effect of transmission fluctuations on clustering of Ly$\alpha$ selected galaxies, beginning with equations~(\ref{taurelation}) and (\ref{analbias}) as motivation. Fluctuations in transmission of Ly$\alpha$ radiation through the IGM modify the intrinsic luminosity that corresponds to an observed flux limit $F_0$. This modification of intrinsic luminosity in turn leads to a modification of number counts according to the luminosity function of Ly$\alpha$ emitters. 

The density $n_{\rm Ly\alpha}$ of Ly$\alpha$ emitters that are observed with fluxes greater than $F_0$ [corresponding to an intrinsic luminosity $L_0$ with transmission $\mathcal{T}_0=\exp{(-\tau_0})$] can then be expressed relative to the average $\bar{n}_{\rm Ly\alpha}(>L_0,\rho_0,\Gamma_0)$ as
\begin{eqnarray}
\nonumber
&&\hspace{-5mm}n_{\rm Ly\alpha}(>F_0) = \bar{n}_{\rm Ly\alpha}(>L_0,\rho_0,\Gamma_0)\times\left(  1+b\delta \right) \\
\nonumber
&&\hspace{5mm}+ (\Gamma-\Gamma_0)\left.\frac{\partial\mathcal{T}}{\partial\Gamma}\right|_{\mathcal{T}_0,\Gamma_0}\left.\frac{\partial \bar{n}_{\rm Ly\alpha}}{\partial\mathcal{T}}\right|_{F_0,\mathcal{T}_0} \\
\nonumber
&&\hspace{5mm}+ (\rho-\rho_0)\left.\frac{\partial\mathcal{T}}{\partial\rho}\right|_{\mathcal{T}_0,\rho_0}\left.\frac{\partial \bar{n}_{\rm Ly\alpha}}{\partial \mathcal{T}}\right|_{F_0,\mathcal{T}_0}\\
&&\hspace{5mm}+ (\frac{dv_z}{d(ar_{\rm com})}-H)\left.\frac{\partial\mathcal{T}}{\partial\frac{dv_z}{d(ar_{\rm com})}}\right|_{\mathcal{T}_0,\rho_0}\left.\frac{\partial \bar{n}_{\rm Ly\alpha}}{\partial\mathcal{T}}\right|_{F_0,\mathcal{T}_0}, 
\end{eqnarray}
where $H$ is the Hubble parameter at scale factor $a$,  and $r_{\rm com}$ is a co-moving distance. Here $\bar{n}_{\rm Ly\alpha}(>L_0)$ is the mean number density of Ly$\alpha$ emitters with luminosities greater than $L_0$, and $n_{\rm Ly\alpha}(>F_0)$ is the perturbed number density of Ly$\alpha$ emitters with observed fluxes greater than the corresponding $F_0$. We define the variable $\delta_\Gamma\equiv\Gamma/\Gamma_0-1$ as the fluctuation in the ionising background (discussed further below). We introduce a change of co-ordinates $\frac{dv^\prime_z}{dr_{\rm com}}\equiv\frac{dv_z}{dr_{\rm com}}-H$, and define the symbol $\delta_v\equiv \frac{dv^\prime_z}{dr_{\rm com}}\frac{1}{Ha}$, which represents the fluctuation in line-of-sight velocity. With these we obtain
\begin{eqnarray}
\nonumber
n_{\rm Ly\alpha}(>F_0) &=& \bar{n}_{\rm Ly\alpha}(>L_0,\rho_0,\Gamma_0)\times\left(  1+b\delta \right) \\
\nonumber
&+& \delta_{\Gamma}\left.\frac{\partial(\log{\mathcal{T}})}{\partial\log{\Gamma}}\right|_{\mathcal{T}_0,\Gamma_0}\left.\frac{\partial \bar{n}_{\rm Ly\alpha}}{\partial(\log{\mathcal{T}})}\right|_{F_0,\mathcal{T}_0}\\
\nonumber
&+& \delta\left.\frac{\partial(\log{\mathcal{T}})}{\partial\log{\rho}}\right|_{\mathcal{T}_0,\rho_0}\left.\frac{\partial \bar{n}_{\rm Ly\alpha}}{\partial (\log{\mathcal{T}})}\right|_{F_0,\mathcal{T}_0}\\
&+& \delta_v\left.\frac{\partial(\log{\mathcal{T}})}{\partial\log{(dv_z/dr)}}\right|_{\mathcal{T}_0,\rho_0}\left.\frac{\partial \bar{n}_{\rm Ly\alpha}}{\partial (\log{\mathcal{T}})}\right|_{F_0,\mathcal{T}_0}. 
\end{eqnarray}

Rearranging we get the fluctuation in the number-density of Ly$\alpha$ emitters 
\begin{equation}
\label{fluctuation}
\delta_{\rm Ly\alpha}\equiv\frac{n_{\rm Ly\alpha}}{\bar{n}_{\rm Ly\alpha}}-1=\delta\left( b+C_{\rho}\right)+\delta_{\Gamma}C_{\Gamma} + \delta_{v}C_{v},
\end{equation}
where
\begin{equation}
C_{\rho}\equiv \frac{1}{\bar{n}_{\rm Ly\alpha}}\left.\frac{\partial \bar{n}(>L)}{\partial (\log{\mathcal{T}})}\right|_{\mathcal{T}_0,L_0}\left.\frac{\partial (\log{\mathcal{T}})}{\partial \log{\rho}}\right|_{\mathcal{T}_0,\rho_0}, 
\end{equation}
\begin{equation}
C_{\Gamma}\equiv   \frac{1}{\bar{n}_{\rm Ly\alpha}}\left.\frac{\partial \bar{n}(>L)}{\partial (\log{\mathcal{T}})}\right|_{\mathcal{T}_0,L_0}\left.\frac{\partial (\log{\mathcal{T}})}{\partial \log{\Gamma}}\right|_{\mathcal{T}_0,\Gamma_0} 
\end{equation}
and
\begin{equation}
\label{Cv}
C_{v}\equiv   \frac{1}{\bar{n}_{\rm Ly\alpha}}\left.\frac{\partial \bar{n}(>L)}{\partial (\log{\mathcal{T}})}\right|_{\mathcal{T}_0,L_0}\left.\frac{\partial (\log{\mathcal{T}})}{\partial \log{(dv_z/dr)}}\right|_{\mathcal{T}_0,\Gamma_0}. 
\end{equation}
By convolving the overdensity field of sources at points $\bf{x}_0$ with a kernel $\propto\exp{[-(\bf{x}-\bf{x}_0)/\lambda]}/(\bf{x}-\bf{x}_0)^2$, the Fourier component of the ionising radiation field can be written \citep[][]{morales2010}
\begin{equation}
\label{eqGamma}
\delta_{\Gamma}(k) = b\delta_k\frac{\arctan{(k\lambda)}}{k\lambda},
\end{equation}
where $\lambda$ is the ionising photon mean-free-path. In redshift space the fluctuation in the density of Ly-$\alpha$ galaxies is
\begin{equation}
\label{eqRedshift}
\delta^{s}_{\rm Ly\alpha} = \delta_{\rm Ly\alpha} - \frac{dv_z}{dr_{\rm com}}\frac{1}{Ha} = \delta_{\rm Ly\alpha} - \delta_v
\end{equation}
Combining equations (\ref{fluctuation}), (\ref{eqGamma}) and (\ref{eqRedshift}), the Fourier component of the fluctuation in space density of Ly$\alpha$ emitters is then
\begin{equation}
\delta^{\rm s}_{\rm Ly\alpha}(k)=\delta_k\left[b\left(1+C_{\Gamma}\frac{\arctan{(k\lambda)}}{k\lambda}\right)+C_{\rho} + (1-C_v)f\mu^2 \right],
\end{equation}
where\footnote{The quantity f is close to unity at high redshifts, taking values of 0.974, 0.988 and 0.997 at z = 2.5, 3.5 and 5.5.} $f = d\log{\delta}/d\log{(1+z)}$, $\mu$ is the cosine between the wave-number and the line-of-sight, and we have used the relation $\delta_v(k)=-f\mu^2\delta_k$ relating fluctuations in velocity gradient and density. The power-spectrum follows directly, yielding
\begin{eqnarray}
\label{PSmu}
P_{\rm Ly\alpha}(k,\mu)=P(k)\left[ b\left(1+C_{\Gamma}K_\lambda\right)+C_{\rho} + (1-C_v)f\mu^2 \right]^2,
\end{eqnarray}
where $P(k)$ is the mass power-spectrum and we have defined 
\begin{equation}
K_\lambda\equiv\frac{\arctan{(k\lambda)}}{k\lambda}.
\end{equation}
The spherically averaged power-spectrum is 
\begin{eqnarray}
\label{PSsph}
\nonumber
P_{\rm Ly\alpha}^{\rm sph}(k)&=&P(k)\left[ \left(b\left(1+C_{\Gamma}K_\lambda\right)+C_{\rho}\right)^2 \right.\\
\nonumber
&+& \frac{2}{3}\left(b\left(1+C_{\Gamma}K_\lambda\right)+C_{\rho}\right)(1-C_v)f \\
&+& \left.\frac{1}{5}(1-C_v)^2f^2 \right].
\end{eqnarray}
We note that if $C_\Gamma=C_\rho=C_v=0$ (indicating no effect from Ly$\alpha$ transmission) we obtain
\begin{equation}
P_{\rm Ly\alpha}(k,\mu) = P_{\rm m}(k)\left[b+   f\mu^2 \right]^{2},
\end{equation}
and 
\begin{equation}
P_{\rm Ly\alpha}^{\rm sph}(k) = P_{\rm m}(k)\left[b^2 +\frac{2}{3}bf  + \frac{1}{5}f^2 \right],
\end{equation}
as expected in the standard case of galaxy clustering~\citep[][]{kaiser1987}.

\section{Analytic model for the effect of Ly$\alpha$ transmission on clustering}

\label{analmodel}

In this section we present a simple parameterised model for the fluctuations in Ly$\alpha$ transmission, and use it to obtain analytic expressions for the constants $C_\rho$, $C_\Gamma$ and $C_v$, and hence for the power-spectrum. The resulting analytic expression is instructive for elucidating the different important effects. In \S~\ref{detailedmodel} we use previously published detailed models of the Ly$\alpha$ emitter transmission to calculate more physically motivated values for $C_\rho$, $C_\Gamma$ and $C_v$.

\subsection{Simple model for $C_\rho$, $C_\Gamma$ and $C_v$}
\label{sec:simplemod}

We begin by assuming a model in which the intrinsic Ly$\alpha$ line of a galaxy is symmetric about the rest-frame Ly$\alpha$ wavelength. We assume the Ly$\alpha$ emitters to be located in an ionised IGM. We also assume that most absorption occurs in regions that are far enough from the galaxy that the ionisation rate is dominated by the ionising background, rather than by ionising flux associated with star-formation in the Ly$\alpha$ emitting galaxy. In the absence of peculiar velocities in the IGM that cause departure from the Hubble flow, the red side of the line is transmitted through the IGM while the blue side is subject to resonant absorption \citep[e.g.][]{Madau95,EHu04}. However infall of intergalactic gas onto massive galaxies leads to resonant absorption that can reach into the red side of the intrinsic line profile \citep[e.g.][see \S~\ref{infall}]{DijkstraIGM,laursen2011}. On the other hand, galactic outflows have the effect of redshifting the emergent Ly$\alpha$ line relative to the true velocity of the galaxy, which makes the Ly$\alpha$ photons more `immune' to scattering in the IGM (see \S~\ref{outflow}).

To maintain generality we therefore assume that a fraction $F$ of the line is subject to absorption in the IGM. As illustrated by equation~(\ref{taurelation}), the transmission of Ly$\alpha$ flux is dependent on the overdensity $\delta$, and on the ionising background and velocity gradient fluctuations $\delta_\Gamma$ and $\delta_v$. The remaining fraction $(1-F)$ is assumed to reach the observer. We therefore have the following expression for the overall transmission of the Ly$\alpha$ line
\begin{equation}
\mathcal{T}(\delta,\delta_\Gamma,\delta_v)=(1-F)+ F\exp{\left({-\tau_0\frac{1+\delta(2.7-0.7\gamma)}{1+\delta_\Gamma+\delta_v}}\right)},
\end{equation}
where $\tau_0$ is the mean Ly$\alpha$ optical depth in the IGM as a whole. This model can describe a wide range of physical scenarios. The case in which the IGM suppresses only the blue half of the line corresponds to $F=0.5$, while infalling intergalactic gas can result in $F>0.5$. Conversely, scattering through galactic winds can cause $F<0.5$.

To calculate $C_\rho$, $C_\Gamma$ and $C_v$ we first need an expression for $d\bar{n}_{\rm Ly\alpha}(>L)/d\log{\mathcal{T}}$. For this calculation we assume that the luminosity function can be approximated as a power-law in the range of luminosities ($L$) observed, 
\begin{equation}
\bar{n}_{\rm Ly\alpha}(>L) =  \bar{n}_{\rm Ly\alpha,0}\left(\frac{L}{L_0}\right)^{1-\beta}.
\end{equation}
We then obtain
\begin{eqnarray}
\nonumber
\left.\frac{d\bar{n}(>L)}{d\log{\mathcal{T}}}\right|_{L_0,\mathcal{T}_0}&=&\left.\frac{d\bar{n}(>L)}{d\log{L}}\right|_{L_0} \left.\frac{d\log{L}}{d\log{\mathcal{T}}}\right|_{L_0,\mathcal{T}_0}\\
&=&(\beta-1)\bar{n}_{\rm Ly\alpha,0},
\end{eqnarray}
where we have used the relation $L\mathcal{T}=L_0\mathcal{T}_0$ (i.e. constant observed flux $F_0$).
Utilising 
\begin{equation}
\left.\frac{d\log{\mathcal{T}}}{d\log{\rho}}\right|_{\mathcal{T}_0,\rho_0} = \frac{-F(2.7-0.7\gamma)\tau_0 e^{-\tau_0}}{(1-F)+Fe^{-\tau_0}}
\end{equation}
and
\begin{equation}
\left.\frac{d\log{\mathcal{T}}}{d\log{\Gamma}}\right|_{\mathcal{T}_0,\Gamma_0} = \left.\frac{d\log{\mathcal{T}}}{d\log{(dv_z/dr)}}\right|_{\mathcal{T}_0,\Gamma_0} = \frac{F\tau_0 e^{-\tau_0}}{(1-F)+Fe^{-\tau_0}},
\end{equation}
we obtain
\begin{equation}
C_\rho = (\beta-1)\frac{(0.7\gamma-2.7)F\tau_0 e^{-\tau_0}}{(1-F)+Fe^{-\tau_0}} 
\end{equation}
and 
\begin{equation}
C_\Gamma = C_v = (\beta-1)\frac{F\tau_0 e^{-\tau_0}}{(1-F)+Fe^{-\tau_0}}. 
\end{equation}
Putting these pieces together we write down an analytic estimate for the power-spectrum of Ly$\alpha$ selected galaxies
\begin{eqnarray}
\label{eq:PSanal}
\nonumber
&&\hspace{-7mm}P_{\rm Ly\alpha}(k,\mu) = \\
&&\hspace{-1mm}P_{\rm m}(k)\left[b+   f\mu^2  + C \left(bK_\lambda + (0.7\gamma-2.7) - f\mu^2\right)\right]^{2},
\end{eqnarray}
where 
\begin{equation}
C\equiv\frac{(\beta-1)F\tau_0 e^{-\tau_0}}{(1-F)+Fe^{-\tau_0}}.
\end{equation}
We also obtain the spherically averaged power-spectrum
\begin{eqnarray}
\nonumber
P_{\rm Ly\alpha}^{\rm sph}(k) &=& P_{\rm m}(k)\left[\left(b  + C \left(bK_\lambda + (0.7\gamma-2.7)\right)\right)^2  \right. \\
\nonumber
&+&     \frac{2}{3} \left(b  + C \left(bK_\lambda + (0.7\gamma-2.7)\right)\right) (1-C)f  \\
&+&  \left.\frac{1}{5} (1-C)^2f^2 \right].
\end{eqnarray}

\begin{figure*}
\includegraphics[width=17cm]{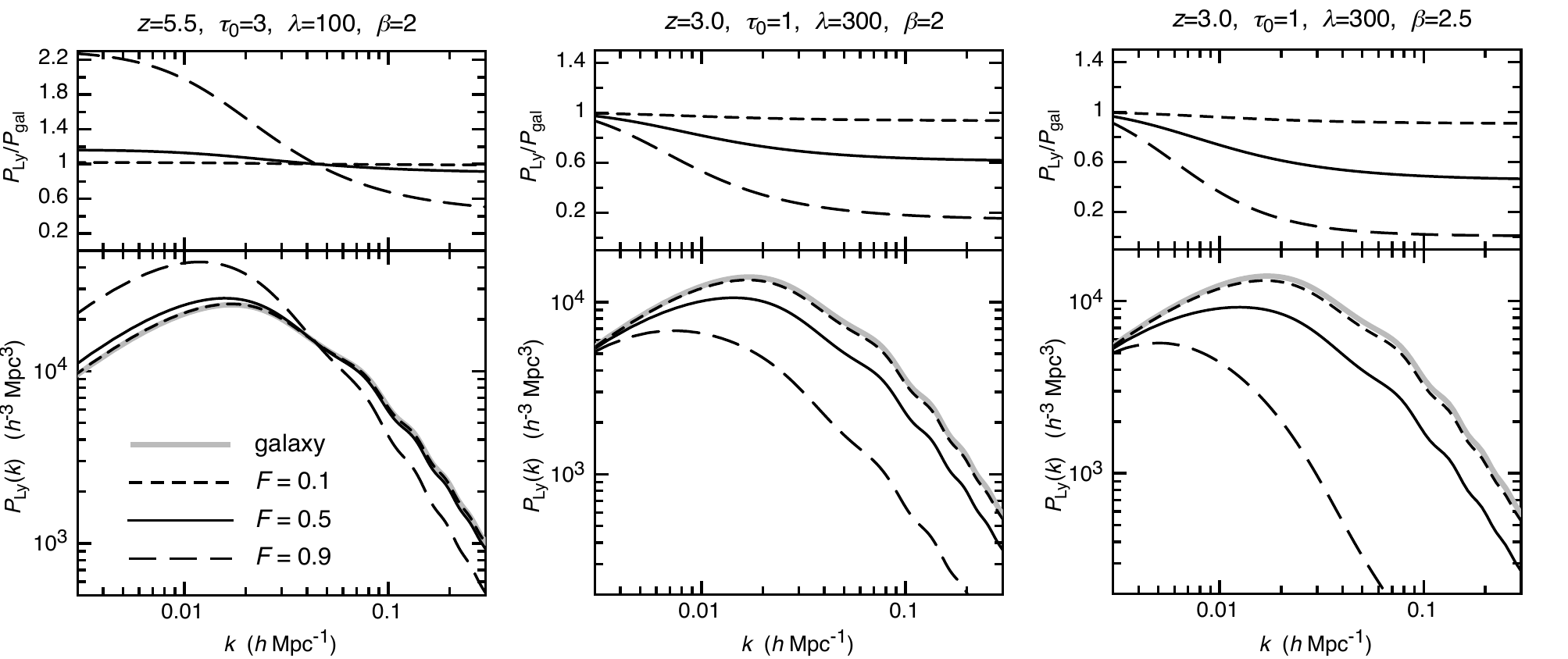}
\caption{Clustering of Ly$\alpha$ emitters in the simple analytic model. The Left, Central and Right panels show the the spherically averaged power-spectra in cases with $(z,\tau_0,\lambda,\beta)=(5.5,3,100$Mpc$,2)$, $(z,\tau_0,\lambda,\beta)=(3.0,1,300$Mpc$,2)$  and $(z,\tau_0,\lambda,\beta)=(3.0,1,300$Mpc$,2.5)$. In each case the short-dashed, solid and long-dashed lines refer to the predicted clustering assuming the analytic model for $C_\rho$, $C_\Gamma$ and $C_v$ with $F=0.1$, 0.5 and 0.9 respectively. The clustering in the absence of transmission effects (i.e. $C_\rho=C_\Gamma=C_v=0$) is shown by the solid grey line. Also shown (upper sub-panels) is the ratio of these curves, indicating the fractional contribution of transmission effects to Ly$\alpha$ emitter clustering. The assumed halo mass was $10^{11}$M$_\odot$. }
\label{fig1}
\end{figure*}

\subsection{Results for analytic model}

Some examples of the predicted power-spectrum of Ly$\alpha$ selected galaxies calculated using this analytic model are shown in Figure~\ref{fig1}. In each case the short-dashed, solid and long-dashed lines refer to the predicted power-spectrum assuming the analytic model for $C_\rho$, $C_\Gamma$ and $C_v$ with $F=0.1$, 0.5 and 0.9 respectively. The power-spectrum in the absence of transmission effects (i.e. $C_\rho=C_\Gamma=C_v=0$) is shown by the solid grey line. The upper sub-panels in each case show the ratio of these curves, indicating the fractional contribution of transmission effects to Ly$\alpha$ emitter clustering. 
Three cases are presented in Figure~\ref{fig1}. In the left column we show a source redshift of $z=5.5$, and assume $\tau_0=3$ and $\lambda=100$cMpc \citep[][]{bolton2007}, with a luminosity function slope\footnote{The faint end slope of the Ly$\alpha$ luminosity function is not well constrained, and may shallower than $\beta=2$ \citep{Ouchi08}. However the LAEs that will be used in the HETDEX survey will be comparable in luminosity to those in the sample of \citet{Ouchi08}. At these luminosities, the slope of the luminosity function is steeper than at the faint end.} of $\beta=2$. In the central column we show results for $z=3.0$, $\tau_0=1$ and $\lambda=300$cMpc \citep[][]{bolton2007}, using the same luminosity function slope of $\beta=2$.   In the right hand column we again show results for $z=3.0$, $\tau_0=1$ and $\lambda=300$cMpc, but this time assume a steep luminosity function slope of $\beta=2.5$.  At each redshift the simple analytic model predicts that fluctuations in transmission can lead to a factor of 2 difference or more in the power-spectrum amplitude. The effects are enhanced by a steep luminosity function, owing to the proportionality of the coefficients $C_\rho$, $C_\Gamma$ and $C_v$ to the factor $(\beta-1)$.

\subsection{Contributions to the clustering amplitude}

\label{contributions}

\begin{figure}
\includegraphics[width=8cm]{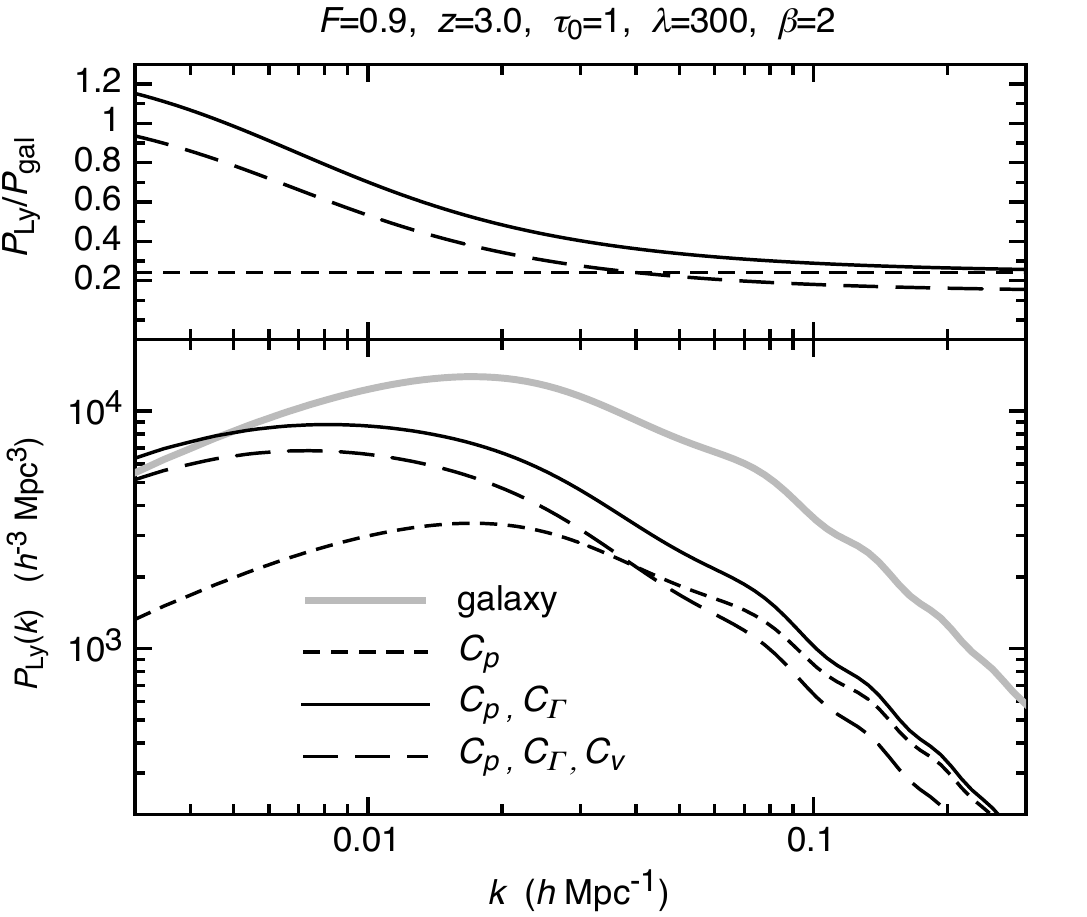}
\caption{Contributions to the spherically averaged clustering of Ly$\alpha$ emitters in the simple analytic model. The case of $z=3.0$, $\tau_0=1$ and $\lambda=300$cMpc, with $\beta=2$ is shown. The short-dashed, solid and long-dashed lines refer to the predicted clustering assuming the analytic model where contributions are included from $C_\rho$, from $C_\rho$ and $C_\Gamma$, and fro all of $C_\rho$, $C_\Gamma$ and $C_v$. We assumed $F=0.9$ to accentuate the dependancies. The clustering in the absence of transmission effects (i.e. $C_\rho=C_\Gamma=C_v=0$) is shown by the solid grey line. Also shown (upper sub-panels) is the ratio of these curves, indicating the fractional contribution of transmission effects to Ly$\alpha$ emitter clustering. The assumed halo mass was $10^{11}$M$_\odot$. }
\label{fig2}
\end{figure}

We next calculate the different contributions to the power-spectrum of Ly$\alpha$ selected galaxies using the analytic model. The case of $z=3.0$, $\tau_0=1$ and $\lambda=300$cMpc, with luminosity function slope $\beta=2$ is shown in Figure~\ref{fig2}. The short-dashed, solid, and long-dashed lines refer to the predicted power-spectrum assuming the analytic model in cases where contributions are included from $C_\rho$, from $C_\rho$ and $C_\Gamma$, and from all of $C_\rho$, $C_\Gamma$ and $C_v$ respectively. We assumed $F=0.9$ to accentuate the dependancies. The clustering in the absence of transmission effects (i.e. $C_\rho=C_\Gamma=C_v=0$) is shown by the solid grey line. 

The simplest effect to understand is the decrease in the power-spectrum amplitude that follows the inclusion of density fluctuations in the Ly$\alpha$ transmission. The density fluctuations preferentially suppress observed flux in overdense regions, leading to lower observed galaxy number densities, and hence to a contribution to the clustering that counteracts galaxy bias. Similarly, the inclusion of fluctuations in velocity gradient also decreases the power-spectrum amplitude. This is because fluctuations in the velocity gradient, which lead to increased transmission and hence to increased galaxy density at fixed Ly$\alpha$ flux, are negatively correlated with over-density in mass.  

\begin{figure*}
\includegraphics[width=17cm]{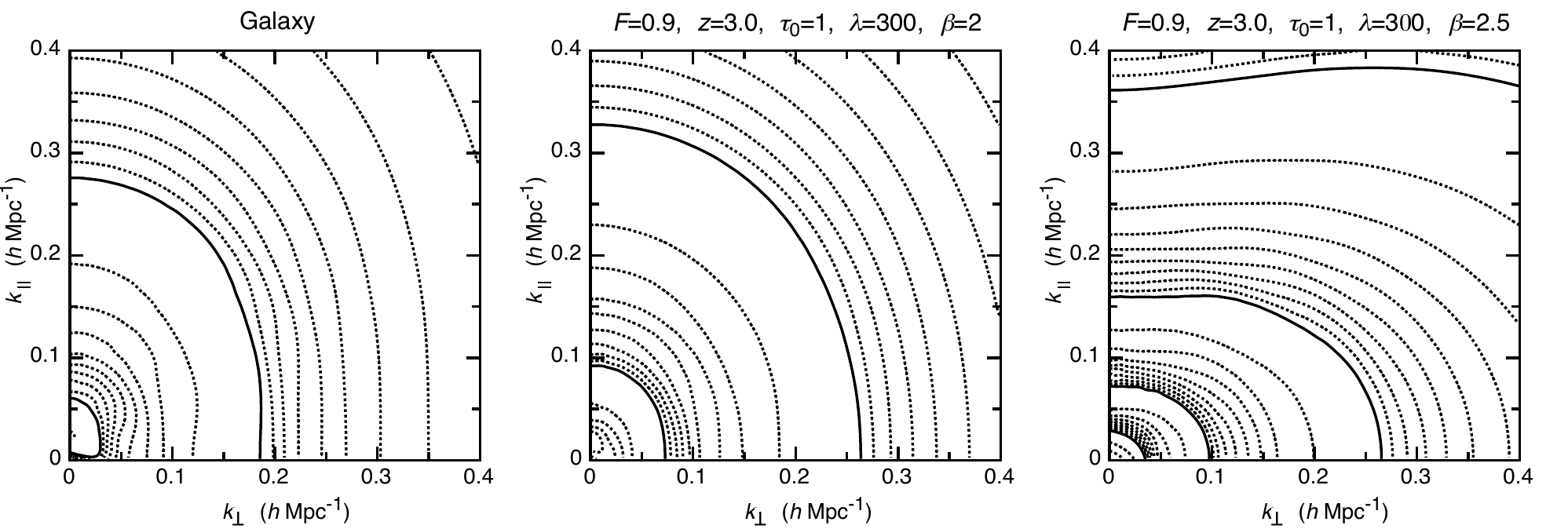}
\caption{ Redshift space clustering of Ly$\alpha$ emitters in the analytic model. The Central and Right panels show contours of the power-spectrum for the cases of $(z,\tau_0,\lambda,\beta)=(3.0,1,300$Mpc$,2)$ and $(z,\tau_0,\lambda,\beta)=(3.0,1,300$Mpc$,2.5)$. The Left panel shows the clustering in the absence of transmission effects (i.e. $C_\rho=C_\Gamma=C_v=0$) for comparison.  The solid contours are separated by a decade in $P_{\rm Ly}$. We have assumed $F=0.9$, and a halo mass of $10^{11}$M$_\odot$.}
\label{fig3}
\end{figure*} 

On large scales ($k\lambda\ll1$) the suppression of the power-spectrum by density fluctuations is counteracted by fluctuations in the ionising background. This is because the fluctuations in ionising background are both correlated with overdensity and biased, and so enhance transmission in overdense regions. In each example shown in Figure~\ref{fig1}, the change of shape in the power-spectrum is evident at the scale $k\sim\pi/(2\lambda)$ corresponding to the ionising photon mean-free-path. On scales smaller than the mean-free-path ($k\lambda\gg1$), fluctuations in the ionising background are washed out and so do not contribute to modification of the clustering amplitude. This is because only a fraction of the ionising background is produced locally within the fluctuation, with the remainder being generated within a larger region that averages over many fluctuations.  On these scales the fluctuations in mass-density result in a lowering of the clustering amplitude by correlating lower than average Ly$\alpha$ transmission with overdensities of galaxies.  It should be noted that our formulation ignores the possible Poisson contribution to fluctuations in the ionising background due to quasars.  Poisson fluctuations introduce additional power beyond the component associated with the underlying density field of galaxies.  The effect of Poisson fluctuations could become important at low redshift ($z\la3$), where quasars contribute significantly to the ionising background but is not included here.

\subsection{Redshift space clustering Ly$\alpha$ galaxies}

In this sub-section we discuss the effect of the velocity structure in the IGM on the observed 2-dimensional redshift space clustering of Ly$\alpha$ selected galaxies on large scales. In Figure~\ref{fig3} we plot contours of the redshift space power-spectrum as a function of the line-of-sight ($k_\parallel=k\mu$) and transverse ($k_\perp=k\sqrt{1-\mu^2}$) components of the wave-number $k$.  The left panel shows the clustering in the absence of transmission effects (i.e. $C_\rho=C_\Gamma=C_v=0$). The effect of infall in the linear regime \citep[][]{kaiser1987} is clearly seen at small values of $k$, resulting in a power-spectrum amplitude that is increased at large scales in the line-of-sight direction. 

The central panel shows contours of the power-spectrum for the cases of $(z,\tau_0,\lambda,\beta)=(3.0,1,300$Mpc$,2)$. We have assumed $F=0.9$ and $M=10^{11}$M$_\odot$. The non-zero $C_v$ term in this model counteracts the \citet[][]{kaiser1987} effect, leading to more isotropic redshift space clustering on large scales. In the right panel we show a more extreme model for the luminosity function with $(z,\tau_0,\lambda,\beta)=(3.0,1,300$Mpc$,2.5)$. In this case  $C_v>1$, and so the \citet[][]{kaiser1987} effect is reversed, leading to suppressed clustering transverse to the line-of-sight \cite[][]{zheng2010}.

\begin{figure*}
\includegraphics[width=17cm]{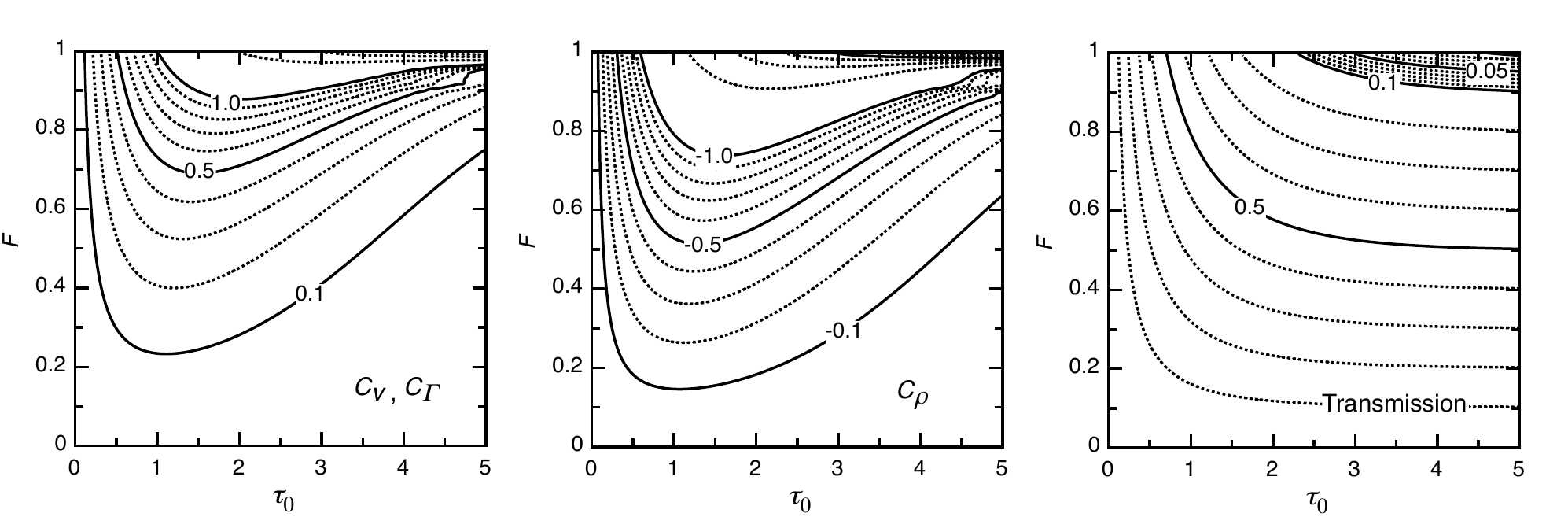}
\caption{ Contours of the coefficients $C_\Gamma$ and $C_v$ (left panel), $C_\rho$ (central panel) and mean transmission $\mathcal{T}_0$ (right panel), calculated as a function of $F$ and $\tau_0$ in our analytic model. We assumed $\beta=2$ for these calculations.}
\label{fig4}
\end{figure*}

\subsection{Coefficients in the Analytic Model}
\label{coefficients}

The coefficients in the analytic model are dependent on assumed values for $F$ and $\tau$. In Figure~\ref{fig4} we plot contours of the coefficients $C_\Gamma$ and $C_v$ (left panel), $C_\rho$ (central panel) and the mean transmission $\mathcal{T}_0$ (right panel), each calculated as a function of $F$ and $\tau_0$. Large values of the coefficients require large values of $F$, indicating that a significant fraction of the intrinsic Ly$\alpha$ line must be subject to absorption in order to influence the clustering of Ly$\alpha$ selected galaxies at a level that is of order unity. 

It is interesting to ask what properties of the model are required to obtain clustering that is enhanced transverse to the line-of-sight as reported in the numerical simulations of \citet[][]{zheng2010}. Inspection of equation~(\ref{eq:PSanal}) indicates that $C_v>1$ is required. Assuming a luminosity function with $\beta=2$, this can be achieved with $F\ga0.9$ and $\tau_0\sim2-5$, for which the transmission is $\mathcal{T}_0\sim 5-20\%$, in good agreement with the Ly$\alpha$ emitters discussed in \citet[][]{zheng2010}. If the luminosity function is steeper, then the requirements on $F$ are less stringent. For $\beta=2.5$, the coefficients are increased by a factor of $(\beta-1)=1.5$ so that  $C_v>1$ is obtained for $F>0.8$.

\section{Detailed modelling of transmission and clustering of Ly$\alpha$ emitters}
\label{detailedmodel}

The analytic model discussed in \S~\ref{analmodel} is useful for investigating the qualitative dependencies of clustering in Ly$\alpha$ selected galaxies. However a more detailed analyses is required to quantitatively predict the values of constants $C_\rho$, $C_\Gamma$ and $C_v$, which describe the modification of the power-spectrum from that measured by a traditional galaxy redshift survey. In this section we describe calculation of $C_\rho$, $C_\Gamma$ and $C_v$ based on two previously published models of Ly$\alpha$ emission. These models explore respectively, the effects of local star-formation and IGM infall \citep[][]{DijkstraIGM}, and of galactic wind driven outflows \citep[][]{verhamme2008,dijkstra2010}, on the transmission of the Ly$\alpha$ line through the circumgalactic IGM.

\subsection{Modelling Ly$\alpha$ transmission in the presence of IGM infall and galactic ionising flux}
\label{infall}

The model presented in this section is based on the work in \citet[][]{DijkstraIGM}, to which the reader is referred for a full description of the calculations. The IGM transmission is calculated using a model for the IGM that accounts for clumping and infall. In this model, resonant absorption of Ly$\alpha$ photons by gas in the infall region (which extends out to several virial radii, see Barkana 2004) erases a significant fraction of the Ly$\alpha$ line flux at frequencies redward of the Ly$\alpha$ resonance. Here, we briefly summarise the modelÕs main ingredients. 

The Ly$\alpha$ flux from star-forming galaxies originates in the dense nebulae from which the stars form. Approximately two out of three ($0.68$) ionising photons produced by O stars (which are absorbed in the nebulae) are converted into Ly$\alpha$ for case-B recombination \citep{Osterbrock89}. The total intrinsic Ly$\alpha$ luminosity of a galaxy can then be estimated from 
\begin{equation}
\label{tran1}
L_{\rm Ly\alpha} = 0.68h\nu_\alpha(1-f_{\rm esc})\dot{Q}_{\rm H},
\end{equation}
where $h\nu_\alpha = 10.2$ eV is the energy of a Ly$\alpha$ photon,  $\dot{Q}_{\rm H}$ is total luminosity of ionising photons, and $f_{\rm esc}$ is the escape fraction of ionising photons from the galaxy. Since $\dot{Q}_{\rm H}$ depends on the number of O-stars, its value is sensitive to the assumed initial mass function (IMF) and metallicity of the gas from which the stars form. Under the assumption of constant star formation rate, \citet[][]{schaerer2003} has calculated $\dot{Q}_{\rm H}$ for several different IMFs, and for a range of metallicities $Z$ (expressed in solar units $Z_\odot$). For a Salpeter IMF with lower and upper mass limits of $M_{\rm l} = 1$M$_\odot$ and $M_{\rm u} = 100$M$_\odot$ respectively
\begin{equation}
\label{tran2}
\log_{10}(\dot{Q}_{\rm H}) = 53.8 + \log_{10}(\dot{M}_\star)-0.0029(9 + \log_{10}(Z))^{2.5},
\end{equation}
in which $\dot{M}_\star$ is the star-formation rate in M$_\odot\,$yr$^{-1}$. We assume $Z = 0.05$. For the models in this sub-section the shape of the intrinsic Ly$\alpha$ line is assumed to be Gaussian (see \S~\ref{outflow} for a discussion of models where we assume different intrinsic spectral line shapes). For gas that is optically thin to Ly$\alpha$ photons, a reasonable choice for the standard deviation of this Gaussian emission line is $\sigma_\alpha\sim v_{\rm vir}$, where $v_{\rm vir}$ is the virial velocity of the host galaxy halo \citep[][]{santos2004,DijkstraIGM}.  We use equation~({\ref{tran2}) to calculate $\dot{Q}_{\rm H}$ as a function of $\dot{M}_\star$. 

\begin{figure*}
\includegraphics[width=17cm]{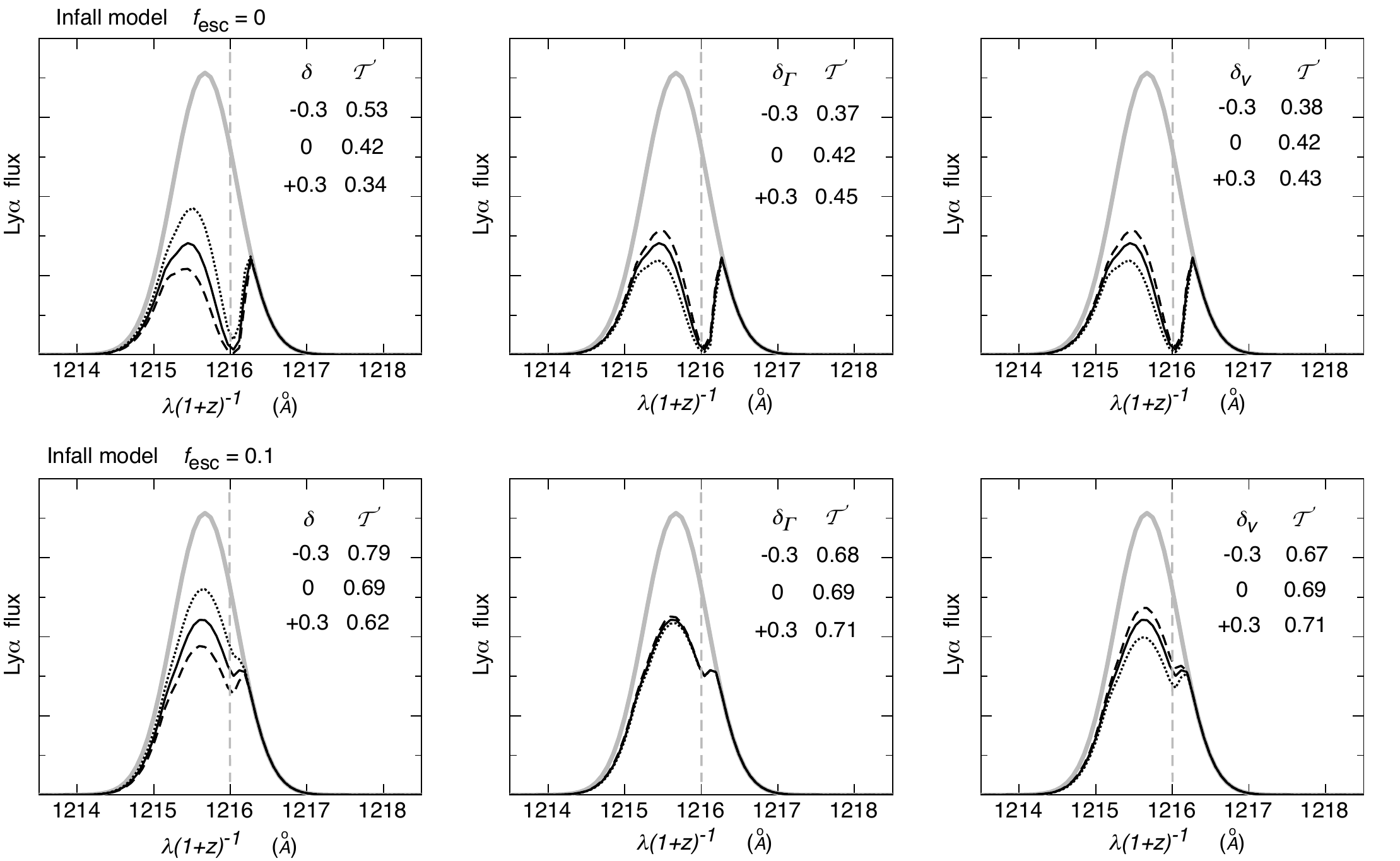}
\caption{Example line profiles in the infall model, with modifications from adjustments in the ionising background $\Gamma$ (left), the density $\rho$ (center), and velocity gradient $dv_z/dr$ (right). The vertical dashed line indicates the Ly$\alpha$ line center. Values for the sizes of the fluctuations considered in these quantities are listed in each case, together with the resulting transmission. The upper row shows results assuming that the galaxy does not contribute to the ionising flux. The lower row assumes an escape fraction of ionising photons from the galaxy of $f_{\rm esc}=10\%$. The assumed halo mass was $10^{11}$M$_\odot$.}
\label{fig5}
\end{figure*}

Given a galaxy spectrum blueward of the hydrogen ionisation  threshold ($\nu_{\rm H}$) of $J(\nu)\propto\nu^{\beta_{\rm s}}$, the photoionisation rate at distance $r$ from the galaxy can be found from
\begin{equation}
\label{eq:gamma}
\Gamma(r) = \Gamma_{\rm bg} + \frac{\beta_{\rm s}}{\beta_{\rm s}-3} \frac{ \sigma_0 \dot{Q}_{\rm H}f_{\rm esc}}{4\pi r^2},
\end{equation}
where $\Gamma_{\rm bg}$ is the photoionisation rate of the meta-galactic background, and we have approximated the hydrogen photoionisation cross-section as $\sigma_H(\nu) = \sigma_0(\nu/\nu_{\rm H})^{-3}$, with $\sigma_0 = 6.3\times10^{-18}$cm$^{-2}$. In addition to the radial dependence of the photoionisation rate we also require the radial density ($\rho$) and velocity ($v$) profiles of the intergalactic gas surrounding objects of mass $M$ \citep[][]{barkana2004}. Useful fitting formula \citep[][]{DijkstraIGM} for these are
\begin{equation}
\label{eq:rho}
\rho(r)=
\left\{ \begin{array}{ll}
         \ 20\bar{\rho}(r/r_{\rm vir})^{-1}& r < 10r_{\rm vir}\\

         \ \bar{\rho} & r \geq 10r_{\rm vir},\end{array} \right.
\end{equation}
and
\begin{equation}
v_{\rm infall}(r)=
\left\{ \begin{array}{ll}
         \ -v_{\rm circ}+ \frac{dv_{\rm infall}}{dr}(r-r_{\rm vir})& r_{\rm vir} <r < 10r_{\rm vir}\\
         \ H(z)r & r \geq 10r_{\rm vir},\end{array} \right.
\label{eq:vel} 
\end{equation} where the velocity gradient $\frac{dv_{\rm infall}}{dr}=(10r_{\rm vir}H(z) +v_{\rm circ})/9r_{\rm vir}$ was chosen to make $v(r)$ continuous. We evaluate the optical depth in this model using equation~(\ref{tranint}), which is also integrated over the probability distribution of density contrasts \citep[][]{miraldaescude2000}. The radial dependencies of $n_H$, $v$ and $\Gamma$ in equation~(\ref{tranint}) are specified by equations~(\ref{eq:gamma}-\ref{eq:vel}). 

For our calculations, we assume an IGM temperature of $T=2\times10^4$ K \citep[e.g.][]{Lidz10}, consider a halo mass of $M=10^{11}$M$_\odot$ \citep[][]{orsi2008,guaita2010}, and a star-formation rate of  $\dot{M}_\star=10$M$_\odot$yr$^{-1}$, which corresponds to the mean and median UV derived SFRs for LAEs in the HETDEX Pilot survey (Blanc et al. 2011). The resulting total intrinsic luminosity in Ly$\alpha$ photons is $L_{\rm Ly\alpha}=2.3\times10^{43}[1-f_{\rm esc}]$ erg/s. The escape fraction of ionizing photons, $f_{\rm esc}$, is uncertain \citep[e.g.][and references therein]{Yajima10}, and may vary significantly between individual objects \citep{Shapley06}. We consider two cases in the paper\footnote{The assumed $f_{\rm esc}$ is degenerate with the assumed gas metallicity $Z$. }: $f_{\rm esc}=0.0$, and $f_{\rm esc}=0.1$. We calculate profiles at $z=3.0$, for which $v_{\rm circ}=103$ km/s and $r_{\rm vir}=39$ kpc). The background photoionisation rate at $z=3.0$ is taken to be $\Gamma_{\rm bg}=0.5\times10^{-12}$ s$^{-1}$ \citep[][]{faucher-giguere2008}. 

Examples of the intrinsic (grey lines) and transmitted (black lines) Ly$\alpha$ lines in this model are shown as a function of restframe wavelength in Figure~\ref{fig5}. We show cases in which no ionising radiation escapes the galaxy (upper panels), and in which the escape fraction is $f_{\rm esc}=0.1$ (lower panels). The line profiles show the features of absorption redward of the intrinsic Ly$\alpha$ central wavelength owing to the influence of gas infall, and no absorption redward of the blue most edge.  The three sets of panels show the dependence of the line profile on fluctuations in the density ($\rho$, left panels), ionising background ($\Gamma_{\rm bg}$, central panels), and velocity gradient ($dv_z/dr$, right panels). For each quantity we present fluctuations (e.g. $\delta\rho\equiv\rho/\rho_0-1$, where $\rho_0$ is the fiducial model) of $\pm0.3$ relative to the fiducial model. The resulting values of transmission $\mathcal{T}$ are presented in Figure~\ref{fig5}. The transmission of the fiducial model is $\mathcal{T}=0.42$ for the case with $f_{\rm esc}=0$, and $\mathcal{T}=0.69$ for the case with $f_{\rm esc}=0.1$.

We incorporate these fluctuations into our model as follows. Firstly, to vary the density $\rho$, we make the modifications $n_{\rm H} \rightarrow n_{\rm H}(1+\delta)$, and ({\it ii}) $T_{\rm gas} \rightarrow T_{\rm gas}(1+[\gamma-1]\delta)$ in equation~(\ref{tauint1}). 
This temperature change affects the recombination coefficient (see equation~\ref{taurelation}). Second, for variations in $\Gamma$, we adjust $\Gamma_{\rm bg}$ in equation~(\ref{eq:gamma}).  Finally, to study the impact of fluctuations  in the velocity gradient, we multiply the optical depth in equation~(\ref{taunu}) by a factor of $(1+\delta_v)$ in the linear regime beyond $10r_{\rm vir}$, and by a factor of ($1+\delta_{\rm infall}$) in the infall region at $r<10r_{\rm vir}$. Though not self-consistent, this procedure preserves the density and velocity profiles, and so isolates the effect of velocity gradient. To evaluate $\delta_{\rm infall}$ we calculate fluctuations in velocity gradient within the infall region relative to the fiducial model with $\left.\frac{dv_{\rm infall}}{dr}\right|_0$. Noting that $\frac{dv_{\rm infall}}{dr}=\frac{10r_{\rm vir} H +v_{\rm circ}}{9 r_{\rm vir}}$ in the infall region, we keep the circular velocity and virial radius fixed, but replace $H$ with $H(1+\delta_v)$ to modify the velocity gradient beyond the infall region.  This results in a fluctuation in the velocity gradient within the infall region of 
\begin{eqnarray}
\nonumber
\delta_{\rm infall} &\equiv& \frac{{dv_{\rm infall}}/{dr}-\left.{dv_{\rm infall}}/{dr}\right|_0}{\left.{dv_{\rm infall}}/{dr}\right|_0}\\
&=&\delta_v\left(\frac{1}{1+10Hr_{\rm vir}/v_{\rm vir}}\right)\approx \delta_v/2,
\end{eqnarray} 
where in the last equality we have noted that the galaxy dynamical time is $r_{\rm vir}/v_{\rm vir}\sim0.1H^{-1}$. Thus fluctuations in velocity gradient within the infall region are reduced relative to those in the linear regime. 

The line profiles in Figure~\ref{fig5} illustrate that fluctuations in density and ionising background lead to substantial modification of the transmitted flux profile on the blue side of the Ly$\alpha$ line, extending into the red side owing to infall. Fluctuations in density have a larger effect on the transmission than fluctuations in ionising background, and the effects have opposite sign as discussed in \S~\ref{contributions}. In the case where there is no contribution to the ionisation of the local IGM from the galaxy ($f_{\rm esc}=0$), the effects of 
fluctuations in ionising background and velocity gradient are of similar magnitude in this model. However, in the case where galactic ionising flux also effects the ionisation state of the local IGM we find that the influence of fluctuations in ionising background is reduced. This is easy to understand since modification of the ionising background level has little effect on the transmission in regions where the galaxy dominates the ionising flux. 

Based on these absorption profiles we can estimate the quantities $\partial \log\mathcal{T}/\partial \log\rho$, $\partial \log\mathcal{T}/\partial \log\Gamma$ and $\partial \log\mathcal{T}/\partial \log(dv_z/dr)$, and hence the values of the constants $C_\rho$, $C_\Gamma$ and $C_v$ which govern the modification of galaxy clustering.  In the case where the galaxy makes no contribution to the ionising flux ($f_{\rm esc}=0$), we find  $C_\rho=-0.72$, $C_\Gamma=0.32$, and $C_v=0.20$. Inspection of Figure~\ref{fig4} shows that these values are similar to those in our simple analytic model for $F\sim0.65$ (corresponding to a fraction of the red half of the line having been absorbed due to infall), and $\tau\sim2$ (which reproduces the infall model transmission at $\sim3$). In the case where $f_{\rm esc}=0.1$, the value of $C_{\Gamma}=0.05$ is much smaller than for $f_{\rm esc}=0$, owing to the reduced relative importance of the ionising background. Similarly, owing to the higher transmission, the $f_{\rm esc}=0.1$ model also leads to lower values of $C_\rho=-0.39$ and $C_v=0.11$. We can again find approximate agreement when we compare these values of $C_\rho$ and $C_v$ to our simple analytic model (Figure~\ref{fig4}) assuming $F\sim0.65$ and $\tau\sim0.5$ (corresponding to the mean transmission of $\mathcal{T}=0.7$ for this model).

\begin{figure*}
\includegraphics[width=17cm]{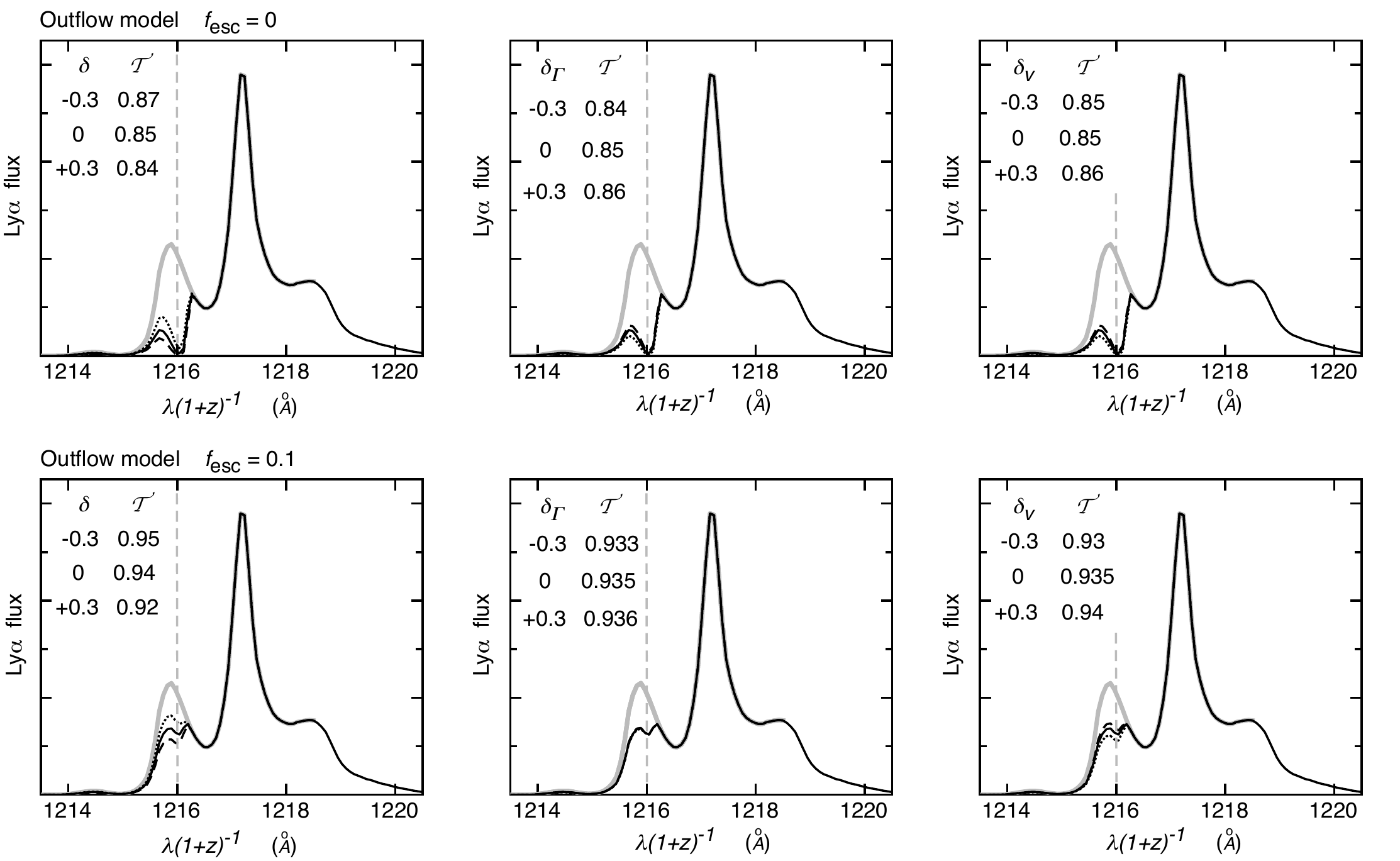}
\caption{Example line profiles in the outflow model, with modifications from adjustments in the ionising background $\Gamma$ (left), the density $\rho$ (center), and velocity gradient $dv_z/dr$ (right). The vertical dashed line indicates the Ly$\alpha$ line center.
Values for the sizes of the fluctuations considered in these quantities are listed in each case, together with the resulting transmission. The upper row shows results assuming that the galaxy does not contribute to the ionising flux. The lower row assumes an escape fraction of ionising photons from the galaxy of $f_{\rm esc}=10\%$. The assumed halo mass was $10^{11}$M$_\odot$.}
\label{fig6}
\end{figure*}

\subsection{ Modelling Ly$\alpha$ transmission in the presence of dust-free, symmetric ISM outflows}
\label{outflow}

The modeling of \S~\ref{infall} assumes the intrinsic Ly$\alpha$ line emerging from the galaxy into the IGM to be a Doppler broadened Gaussian, which is symmetric in frequency around the Ly$\alpha$ resonance. However, galactic outflows have the effect of redshifting the emergent Ly$\alpha$ line relative to the true velocity of the galaxy \citep[e.g.][]{Ahn03,verhamme2006}, and there is strong observational evidence that this mechanism is at work. This evidence includes the observed blue-shift of interstellar metal absorption lines combined with the observed red-shift of the Ly$\alpha$ emission line \citep{Steidel10}, and the fact that Ly$\alpha$ line shapes are asymmetric at all redshifts \citep[e.g.][]{Mas03,Heckman11}. We refer the reader to \citet{dijkstra2011} for a more extended discussion. \citet{verhamme2006,verhamme2008} have developed a simple model in which scattering of Ly$\alpha$ photons by \ion{H}{I} in these outflows successfully explaines the observed Ly$\alpha$ line shapes observed in Ly$\alpha$ emitting galaxies at $z=3-6$ \citep[also see][]{vanzella2010}. 
  
In this section we repeat the exercise of \S~\ref{infall} for a suite of outflow models. Following \citet{verhamme2006,verhamme2008} and \citet[][]{dijkstra2010}, we model the outflow as a spherically symmetric thin shell of gas that contains an \ion{H}{I} column density $N_{\rm HI}$, and outflow velocity $v_{\rm sh}$. We assume that the shell has a radius of $1$ kpc and a thickness of 0.1 kpc, but stress that the precise physical scale of the outflow is not important for our results. Our assumed gas temperature of $T_{\rm ISM}=10^4$ K in the outflowing \ion{H}{I} shell corresponds to a $b$-parameter of $\sim 13$ km s$^{-1}$ in the terminology of  \citet[][]{verhamme2008}. We further assume the \ion{H}{I} shells to be dust-free (see \S~\ref{sec:dust} for a discussion on dusty outflows).  \citet[][]{verhamme2008} typically found that $\log N_{\rm HI}\sim 19-21$, and $v_{\rm sh}\sim 0-500$ km s$^{-1}$.  We therefore assume a model in which $(N_{\rm HI},v_{\rm sh})=(10^{20}\hs{\rm cm}^{-2},200\hs{\rm km} \hs {\rm s}^{-1})$. We compute Ly$\alpha$ spectra emerging from the outflows using a Monte-Carlo transfer code \citep{dijkstra2006}. In our calculations, the Ly$\alpha$ photons are emitted at line center ($\lambda_{\rm Ly\alpha}=1216\AA$). We compute the impact of the IGM on the directly observed fraction of Ly$\alpha$ by suppressing the intrinsic spectrum by exp$(- \tau)$ (see \S~\ref{infall}). Further details on the calculation of this model can be found in \citet[][]{dijkstra2011}. The {\it grey solid line} in Figure~\ref{fig6} shows an example of the Ly$\alpha$ spectra emerging from the outflows. The emerging spectrum is highly asymmetric, with more flux coming out on the red side of the Ly$\alpha$ line center. The spectrum peaks at about $\sim 2v_{\rm sh}$, as expected for radiation that scatters back to the observer on the far side of the galaxy \citep[see][for a detailed discussion on these features in the spectrum]{Ahn03,verhamme2006}. 

In Figure~\ref{fig6} we show the transmitted (black lines) Ly$\alpha$ line for the outflow model. As in Figure~\ref{fig5} we show cases in which no ionising radiation escapes the galaxy (upper panels), and in which the escape fraction is $f_{\rm esc}=0.1$ (lower panels).  As before we show the dependence of the line profile on each of fluctuations in the density ($\rho$ left panels), ionising background ($\Gamma_{\rm bg}$, central panels), and velocity gradient ($dv_z/dr$, right panels). The resulting values of transmission $\mathcal{T}$ are listed. The transmission of the fiducial outflow model is $\mathcal{T}=0.85$ where $f_{\rm esc}=0$, and $\mathcal{T}=0.94$ where $f_{\rm esc}=0.1$. These values are larger than were found for the infall case, owing to the large fraction of radiation that scatters away from resonance before emerging from the galaxy. All trends of the transmission with fluctuations in density, ionising background and velocity gradient previously described for the infall model, are also present in the outflow models. However the modifications of the transmitted flux are smaller than are found in the infall case. This is because a smaller fraction of the emergent flux is subject to resonant absorption in the IGM. 

Based on these absorption profiles we can estimate the values of the constants $C_\rho$, $C_\Gamma$ and $C_v$ which govern the modification of galaxy clustering in the outflow model. In the case where the galaxy makes no contribution to the ionising flux, we find  $C_\rho=-0.06$,  $C_\Gamma=0.04$ and $C_v=0.02$. Inspection of Figure~\ref{fig4} shows these values are similar to our simple analytic model where $F\sim0.2$ (corresponding to most of the line having been scattered redward of Ly$\alpha$ by the outflow), and $\tau\sim2$ (which results in a transmission equal to that predicted by the detailed outflow model). In the case where $f_{\rm esc}=0.1$, the value of $C_\Gamma=0.005$ is much smaller than for the $f_{\rm esc}=0$ case owing to the reduced importance of the ionising background. Similarly, because of the higher transmission, the model with  $f_{\rm esc}=0.1$ also leads to lower values of $C_\rho=-0.05$ and $C_v=0.01$. We again find approximate agreement when we compare these values for $C_\rho$ and $C_v$ to our simple analytic model (Figure~\ref{fig4}) with $F\sim0.2$ and $\tau\sim0.5$, corresponding to the mean transmission of $\mathcal{T}=0.94$ for this model. 

\subsection{Modeling the IGM transmission for dusty, anisotropic outflows}
\label{sec:dust}

The calculations presented in \S~\ref{outflow}  ignore the impact of dust on the Ly$\alpha$ radiation field. However dust can play an important role in the scattering of Ly$\alpha$ photons within galaxies \citep[e.g.][]{Neufeld91,Hansen06}. \citet{Laursen09} have performed Ly$\alpha$ radiative transfer calculations in simulated galaxies, finding that the effect of dust is to narrow the Ly$\alpha$ line emerging from a galaxy relative to the dust-free case\footnote{This is mostly because Ly$\alpha$ scattering can be described by a diffusion process in both real and frequency space \cite[see][and references therein]{dijkstra2006}. A large displacement in frequency requires a large number of scatterings, and therefore a long trajectory through the scattering medium. As the dust content of this medium is increased, the probability that the photon is destroyed by a dust grain is enhanced.}. Narrowing the Ly$\alpha$ line causes a larger fraction of Ly$\alpha$ photons to emerge at frequencies where they are subject to scattering in the IGM. We therefore expect the impact of the IGM to be stronger in cases where dust is included in the modelling of outflows. 

In our model, the galactic outflow is represented by a spherical shell. Departures from this idealized gas distribution should also lead to a larger impact of the IGM on the emerging Ly$\alpha$ line. This is because Ly$\alpha$ photons will escape more easily from outflows in which either the covering factor is less than unity (because some sight-lines simply do not intersect with the outflowing material), or when the scattering medium is clumpy. The latter is demonstrated by \citet{Hansen06} who have studied Ly$\alpha$ transfer through clumpy outflows, and have shown that a fraction of the photons can escape at line center. These results suggest that more complicated, and realistic models of winds than those employed in this paper will result in a stronger impact of the IGM on the observed Ly$\alpha$ flux than we have computed here \citep[\S~\ref{outflow}, also see][]{Barnes11}.

\subsection{Summary of detailed modelling}
 
The values of $C_\rho$, $C_\Gamma$ and $C_v$ found from the detailed modelling in this section are summarised in Table~\ref{tab1}. The prediction from the infall model for Ly$\alpha$ transmission is that that there will be significant contributions to the observed power-spectrum from fluctuations in the Ly$\alpha$ transmission. Indeed, if the escape fraction of ionising photons is very small, then the contributions are expected to be of order unity. In contrast, the prediction from the outflow model is that contributions to the observed power-spectrum will be an order of magnitude smaller, at the level of $\sim5-10\%$, although these numbers are likely to be conservatively small  (see \S~\ref{sec:dust}). 

These results suggest that measurement of the terms $C_\rho$, $C_\Gamma$ and $C_v$ from the observed power-spectrum of Ly$\alpha$ selected galaxies will provide a new avenue to study the relationship between the Ly$\alpha$ flux of galaxies and their local IGM. On the other hand, the expected non-zero values of $C_\rho$, $C_\Gamma$ and $C_v$ may complicate attempts to use the power-spectrum of Ly$\alpha$ selected galaxies to constrain cosmological parameters. We turn to this topic for the remainder of the paper, in which we present an application of our general model for clustering of Ly$\alpha$ selected galaxies to the planed HETDEX survey. 

\begin{table}
\begin{center}
\caption{\label{tab1} Evaluations of the constants $C_\rho$, $C_\Gamma$ and $C_v$ for the different models.}
\begin{tabular}{cccc}
\hline
 model                                     & $C_\rho$   & $C_\Gamma$ &  $C_v$     \\\hline
 infall  ($f_{\rm esc}=0$)       &        -0.72         &          0.32           &       0.20         \\
  infall  ($f_{\rm esc}=0.1$)    &        -0.39        &             0.05           &     0.11         \\
 outflow  ($f_{\rm esc}=0$)     &       -0.06          &       0.04        &        0.02    \\
  outflow  ($f_{\rm esc}=0.1$)  &       -0.05          &              0.005                &         0.01       \\\hline
\end{tabular}
\end{center}
\end{table}

\section{Ly$\alpha$ transmission fluctuations in galaxy redshift surveys}
\label{HETDEX}

One of the primary science drivers motivating large galaxy redshift surveys is measurement of dark energy,  and its evolution. Traditional galaxy redshift surveys are best suited to studies of the dark energy equation of state at relatively late times ($z \la 1$) due to the difficulty of obtaining accurate redshifts for a sufficiently large number of high redshift galaxies. Although detection of the Integrated Sachs-Wolfe effect puts some constraints on the integrated role of dark energy above $z \sim 1.5$ [see, e.g., \citet{giannantonio2008} and references therein], we currently have very limited information about the nature of dark energy at high redshift. If dark energy behaves like a cosmological constant, then its effect on the Hubble expansion is only significant at $z\la1$ and becomes small at $z\ga2$. In this case, studies of the power-spectrum at low redshift would provide the strongest constraints.  However, as the origin of dark energy is not understood we cannot presume a~priori which redshift range should be studied in order to provide optimal constraints on proposed models. Probes of dark energy at higher redshifts have been suggested. These include measurements of the power-spectrum from a Ly$\alpha$ forest survey, which could potentially be used probe the evolution of dark energy through measurement of the baryonic acoustic oscillation (BAO) scale scale for redshifts as high as $z\sim4$ \citep{mcdonald2007}. Similarly, studying the temporal variation of high resolution quasar spectra may probe the evolution of dark energy in the window $2<z<5$ \citep{corasaniti2007}. 

The Hobby-Eberly Telescope Dark Energy Experiment \citep[][HETDEX]{hill2004b,hill2008a}, promises to provide a very important advance in our understanding of dark energy, by measuring its contribution to the energy density at high redshift ($z\sim2.5$) where there are currently no direct constraints. At the same time, a precision measurement of curvature will assist in breaking the degeneracy between dark energy and curvature present in lower redshift experiments \citep[][]{hill2004b,mcdonald2007}. The HETDEX approach is to obtain approximately 0.8 million redshifts of Ly$\alpha$ selected galaxies at $1.9 < z < 3.5$. These galaxies will be obtained over an area of 400 square degrees, with a survey volume of $V\sim9$Gpc$^{3}$, and a galaxy space density of $n_{\rm Ly}\sim10^{-4}$Mpc$^{-3}$. In the absence of non-gravitational contributions to the clustering, such a survey is able to use the measured power-spectrum to determine the local Hubble expansion at $z\sim2.5$, and the angular diameter distance out to $z\sim2.5$ to 0.8\% each. In this section we discuss the influence that the non-gravitational contribution to the observed power-spectrum of Ly$\alpha$ selected galaxies may have on the precision of cosmological constraints from a survey like HETDEX. We find that the transmission can strongly influence the constraints that are available, and determine the precision with which the parameters $C_v$, $C_\Gamma$ and $C_\rho$ will need to be understood in order for HETDEX to achieve its theoretical precision.

\subsection{Power-spectrum and power-spectrum sensitivity}

\begin{figure*}
\includegraphics[width=17cm]{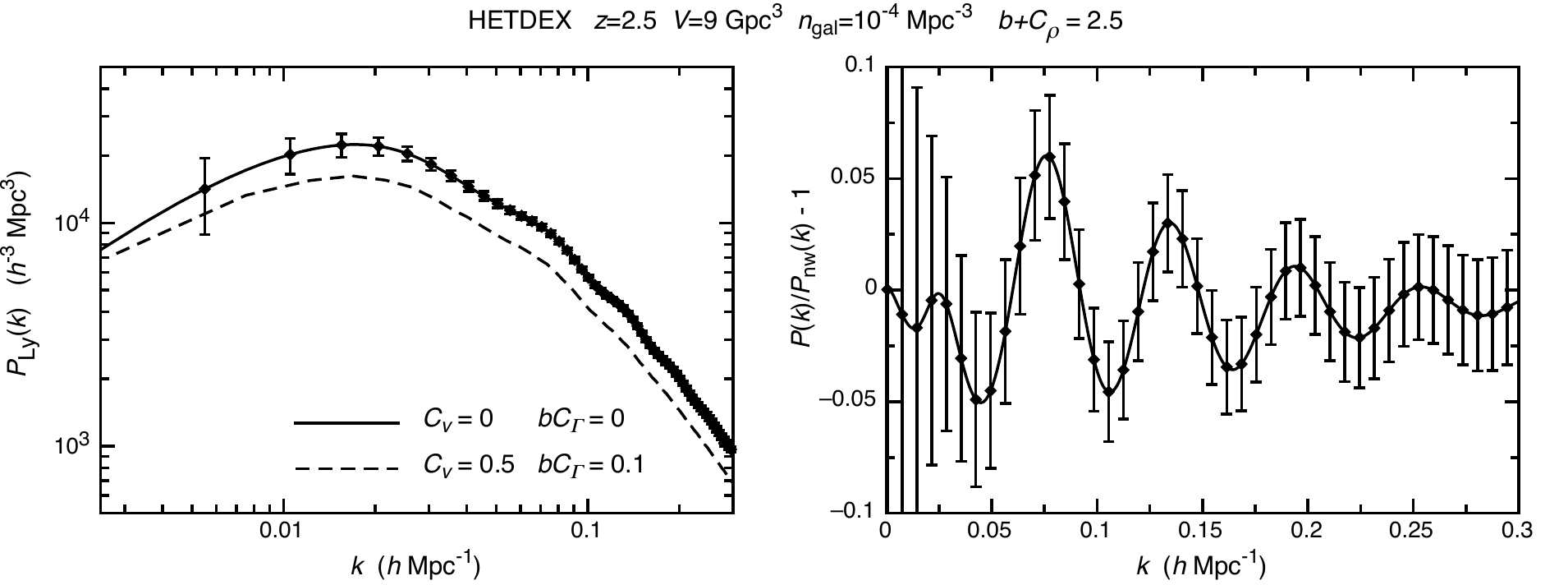}
\caption{{\em Left panel:} The predicted spherically averaged power-spectrum of Ly$\alpha$ galaxies, using equation~(\ref{PSsph}) with the combinations $[C_v,bC_\Gamma,(b+C_\rho)]=(0,0,2.5)$ and $(0.5,0.1,2.5)$. {\em Right panel:} The baryonic acoustic oscillations, plotted as fluctuations relative to the ``no wiggle'' power-spectrum. The error-bars assuming the observational survey parameters $V\sim9$Gpc$^{3}$ and $n_{\rm Ly}\sim10^{-4}$Mpc$^{-3}$ that are appropriate for HETDEX, with each point providing an independent measures of the power. }
\label{fig7}
\end{figure*}

In this section we describe the power-spectrum, and the estimate of power-spectrum sensitivity, that we employ to calculate the influence of transmission fluctuations on cosmological constraints that will be available in HETDEX. Analysis of the galaxy power-spectrum derived from N-body simulations has shown \citep[][]{seo2005} that the power-spectrum can be treated as linear on scales greater than 15 co-moving Mpc (i.e. $k_{\rm max}\lsim0.4$Mpc$^{-1}$) at $z=3.5$, increasing towards higher redshifts. Our assumption of a linear mass power-spectrum should therefore be sufficient for this analysis. However, weak oscillatory features in the power-spectrum, such as the baryonic acoustic oscillations, are suppressed on even larger scales because matter moves across distances on the order of $\sim 5$--10Mpc over a Hubble time\footnote{This characteristic scale of displacement follows from the fact that $\sigma_8$, the normalisation of the power-spectrum on 8$h^{-1}$Mpc, is of order unity at the present time.}. As groups of galaxies form, the linear-theory prediction for the location of each galaxy becomes uncertain, and as a result noise is added to the correlation among galaxies and hence to the measurement of the mass power-spectrum. The noise associated with the movement of galaxies smears out the acoustic peak in the correlation function of galaxies in both real and space \citep[][]{eisenstein2007,seo2008}. The associated reduction of power in the baryonic acoustic oscillations is found to be in excess of 70\% on scales smaller than $k_{\rm max}\sim0.4$Mpc$^{-1}$ at $z\sim3$, corresponding to a length scale of $\sim \pi/(2k_{\rm max})= 3.9$ comoving Mpc \citep[][]{seo2008}. For our cosmological analysis we therefore use the following mass power-spectrum 
\begin{eqnarray}
\nonumber
&&\hspace{-7mm}P_{\rm m,nl}(k)= P_{\rm m,nw}(k) \\
&&\hspace{-5mm}+\left(P_{\rm m}(k) - P_{\rm m,nw}(k)\right) \exp{\left(-k^2\frac{\Sigma_\perp^2(1-\mu^2) + \Sigma_\parallel^2\mu^2}{2}\right)},
\end{eqnarray}
where $P_{\rm m,nw}$ is the ``no wiggle'' form from \citet[][]{eisenstein1999}. The non-linear scales in this expression are $\Sigma_\perp=4.6[(1+z)/3.5]^{-1}$Mpc and $\Sigma_\parallel=9.2[(1+z)/3.5]^{-1}$Mpc \citep[][]{seo2007} in the high redshift limit. Note that we consider only scales $k<0.4$Mpc$^{-1}$, and so do not include the effects of the fingers-of-god that arise from random motions within virialized halos in our analysis. As shown in  \citet[][]{shoji2009}, this has no influence on the cosmological constraints inferred from the large scale power-spectrum.

Like traditional galaxy redshift surveys, the observed Ly$\alpha$ galaxy power-spectrum is sensitive to the underlying mass power-spectrum ($P_{\rm m}$) and galaxy bias $b$. However in addition, there is also dependence on the parameters $C_\Gamma$, $C_\rho$, $C_{v}$, which are related to properties of Ly$\alpha$ transmission through the IGM, and on the mean-free-path $\lambda$. Inspection of equations~(\ref{PSmu}) and (\ref{PSsph}) indicates that not all of $b$, $C_\rho$ and $C_\Gamma$ can be measured independently. Instead, the power-spectrum of Ly$\alpha$ emitting galaxies depends on the parameters $bC_\Gamma$, $(C_\rho+b)$ and $C_{v}$. In the left panel of Figure~\ref{fig7} we plot the predicted spherically averaged power-spectrum of Ly$\alpha$ galaxies, using equation~(\ref{PSsph}) with the combinations $[C_v,bC_\Gamma,(b+C_\rho)]=(0,0,2.5)$ and $(0.5,0.1,2.5)$. These parameters are motivated by an extreme outflow model (\S~\ref{outflow}), and an infall model (\S~\ref{infall}) respectively. 

Figure~\ref{fig7} also presents error-bars assuming the observational survey parameters $V\sim9$Gpc$^{3}$ and $n_{\rm Ly}\sim10^{-4}$Mpc$^{-3}$ appropriate for HETDEX. We evaluate the uncertainty in a k-space volume $2\pi k^2\,\Delta k\,\sin(\mu)\,\Delta\mu$ as
\begin{equation}
\label{PSnoise}
\Delta P_{\rm Ly} = P_{\rm Ly}\left(1+\frac{1}{P_{\rm Ly}n_{\rm Ly}}\right)\left(k^2\,\Delta k\,\sin(\mu)\Delta\mu \frac{V}{(2\pi)^2.}\right)^{-1/2},
\end{equation}
where the sum in the first term encapsulates cosmic variance and galaxy shot-noise respectively, and the second term corresponds to the number of modes measured in the survey. We find that shot-noise dominates at $k\ga0.1$Mpc$^{-1}$. The right panel of Figure~\ref{fig7} shows the baryonic acoustic oscillations, plotted as fluctuations relative to the ``no wiggle'' power-spectrum.

\subsection{Measurement baryonic acoustic oscillations with Ly$\alpha$ selected galaxies }

\begin{figure*}
\includegraphics[width=17cm]{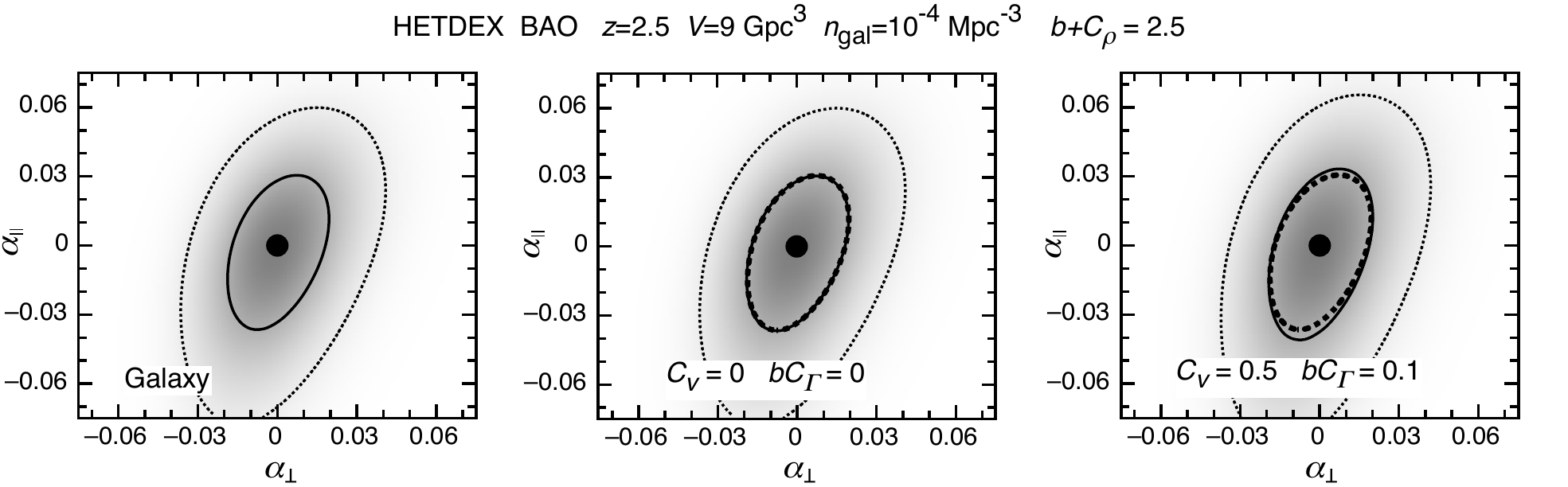}
\caption{Constraints on the line-of-sight and parallel BAO scale. The three panels show contours of likelihood for the parameter set $(\alpha_\perp,\alpha_\parallel)$, in the cases of a traditional galaxy power-spectrum (left), and Ly$\alpha$ galaxy power-spectra with $[C_v,bC_\Gamma,(b+C_\rho)]=(0,0,2.5)$ (central) and $[C_v,bC_\Gamma,(b+C_\rho)]=(0.5,0.1,2.5)$ (right) respectively. Contours of likelihood for the values of $\alpha_\parallel$ and $\alpha_\perp$ are shown (at 61\% and 14\% of  the peak likelihood). The values of $61\% \sim e^{-\Delta \chi^2/2}$ and $14\% \sim e^{-\Delta \chi^2/2}$ are chosen to equal the contour height corresponding to $\Delta \chi^2=1$ and $\Delta \chi^2=4$, and therefore the projection of these contours onto the axis for a particular parameter represents the 1-sigma and 2-sigma ranges respectively. The 61\% contour for the traditional galaxy redshift survey constraints is repeated in the central and right panels for comparison (thick dotted line).}
\label{fig8}
\end{figure*}

The imprint of BAOs on the mass power spectrum provides a cosmic yardstick that can be used to measure the dependence of both the angular diameter distance and Hubble parameter on redshift. The wavelength of a BAO is related to the size of the sound horizon at recombination. Its value depends on the Hubble constant, and on the dark matter and baryon densities. However, it does not depend on the amount or nature of dark energy. Thus measurements of the angular diameter distance and Hubble parameter can in turn be used to constrain the possible evolution of dark energy with cosmic time \citep[e.g.][]{eisenstein1998,eisenstein2002}.

Importantly, a measurement of the BAO scale is not subject to modifications of the overall shape or angular dependence of the power-spectrum, owing to non-linear gravitational growth. To illustrate this point \citet{seo2005} modelled the power-spectra from a series of N-body simulations using the addition of a linear power-spectrum and a scale dependent polynomial to describe galaxy bias and anomalous power. \citet{seo2005} find that they are able to recover the BAO signal by subtracting a smooth function from the matter power-spectrum measured in their N-body simulations. 

Our model is intrinsically linear and so does not include scale dependent bias or anomalous power. However the subtraction of a smooth function to recover the BAO signal is valid for a range of different scale dependent contributions \citep[e.g.][]{mao_tegmark2008,rhook2009}. As a result, while Ly$\alpha$ transmission fluctuations inhibit the extraction of the full cosmological information through consideration of the whole power-spectrum, they should not significantly alter the ability of a Ly$\alpha$ galaxy survey to measure the BAO scale. 

To illustrate this point we have fitted the analytic approximation to the baryonic oscillation component of the redshift space power spectrum following \citet[][]{glazebrook2005}
\begin{eqnarray}
\label{BAOPS}
\nonumber
P_{\rm Ly}(k_\parallel,k_\perp) &=& P_{\rm Ly,nw}(k_\parallel,k_\perp)\\
\nonumber
&&\hspace{-20mm}\times \left\{1+ A k \exp{\left[-\left(\frac{k}{0.07\mbox{Mpc}^{-1}}\right)^{1.4}\right]}\right.\\
\nonumber
&&\hspace{-20mm}\left.\times\sin\left(2\pi\sqrt{\left(\frac{k_\perp}{(1+\alpha_\perp)k_{\rm A}}\right)^2+\left(\frac{k_\parallel}{(1+\alpha_\parallel)k_{\rm A}}\right)^2}\right)\right.\\
&&\hspace{-0mm}\times\left.\exp{\left(-k^2\frac{\Sigma_\perp^2(1-\mu^2) + \Sigma_\parallel^2\mu^2}{2}\right)}\right\},
\end{eqnarray}
to estimate the constraints on the line-of-sight and transverse BAO scales ($\alpha_\parallel$ and $\alpha_\perp$). In this expression the "wiggle free" power-spectrum ($P_{\rm Ly,nw}$) is computed using equation~(\ref{PSmu}), with the mass power-spectrum ($P_{\rm m}$) replaced by the "wiggle-free" mass-power-spectrum ($P_{\rm m,nw}$). The observed power-spectrum $P_{\rm Ly}$ is modelled as the sum of $P_{\rm Ly,nw}$ and a decaying sinusoid with characteristic periods in the line-of-sight and transverse directions of  $(1+\alpha_\parallel)k_{\rm A}$ and $(1+\alpha_\perp)k_{\rm A}$. We include the factor of non-linear suppression of the BAO amplitude \citep[][]{seo2007}. This function has three parameters $A$, $\alpha_\parallel$ and $\alpha_\perp$. The value of $A$ is determined to high accuracy from observations of the Cosmic Microwave Background. For the purposes of this analysis we therefore assume that $A$ is a known constant (namely $A=2.1$), and fit only for $\alpha_\parallel$ and $\alpha_\perp$ (around the best fit value of $k_{\rm A}$ in the absence of noise). We fit only to values of $k<0.4$Mpc$^{-1}$. With this parameterisation, the accuracy with which $\alpha_\parallel$ and $\alpha_\perp$ can be measured determines the constraints that BAO can place on the line-of-sight and transverse distances, and hence on the Hubble parameter $H$ and angular diameter distance $D_{\rm A}$ respectively. 

In Figure~\ref{fig8} we show contours of $\alpha_\parallel$ and $\alpha_\perp$ derived from the BAO analysis. For graphical representation of our results we show contours at 61\% and 14\% of the peak height. The values of $61\% \sim e^{-\Delta \chi^2/2}$ and $14\% \sim e^{-\Delta \chi^2/2}$ are chosen to equal the contour height corresponding to $\Delta \chi^2=1$ and $\Delta \chi^2=4$, and therefore the projection of these contours onto the axis for a particular parameter represents the 1-sigma and 2-sigma ranges respectively. The left hand panel shows the expectations for a galaxy redshift survey with parameters corresponding to HETDEX. Our analysis yields 1-sigma errors of $\Delta \alpha_\parallel\sim3.0\%$ and $\Delta\alpha_\perp\sim2.0\%$. For consistency we compare with the BAO constraints for HETDEX presented in \citet[][]{shoji2009}. These authors find similar constraints of $\Delta \alpha_\parallel\sim2.5\%$ and $\Delta\alpha_\perp\sim1.8\%$. 

In the central and right hand panels of Figure~\ref{fig8} we allow for scale and direction dependent modifications to the power-spectrum owing to fluctuations in Ly$\alpha$ transmission (equation~\ref{PSmu}). In these cases the fiducial models have $(C_v,bC_\Gamma)=(0,0)$ and  $(C_v,bC_\Gamma)=(0.5,0.1)$ respectively. No prior probabilities on their values were assumed. There is a very small difference in the resulting BAO constraints, indicating that modification of the power-spectrum by fluctuations in  Ly$\alpha$ transmission will not inhibit use of the BAO scale for studies of dark energy in Ly$\alpha$ selected galaxy surveys. The values for constraints on $\alpha_\parallel$ and $\alpha_\perp$ are listed in Table~\ref{tab2}.

\subsection{Application of the Alcock-Paczynski test}

\label{secAP}
\begin{figure*}
\includegraphics[width=17cm]{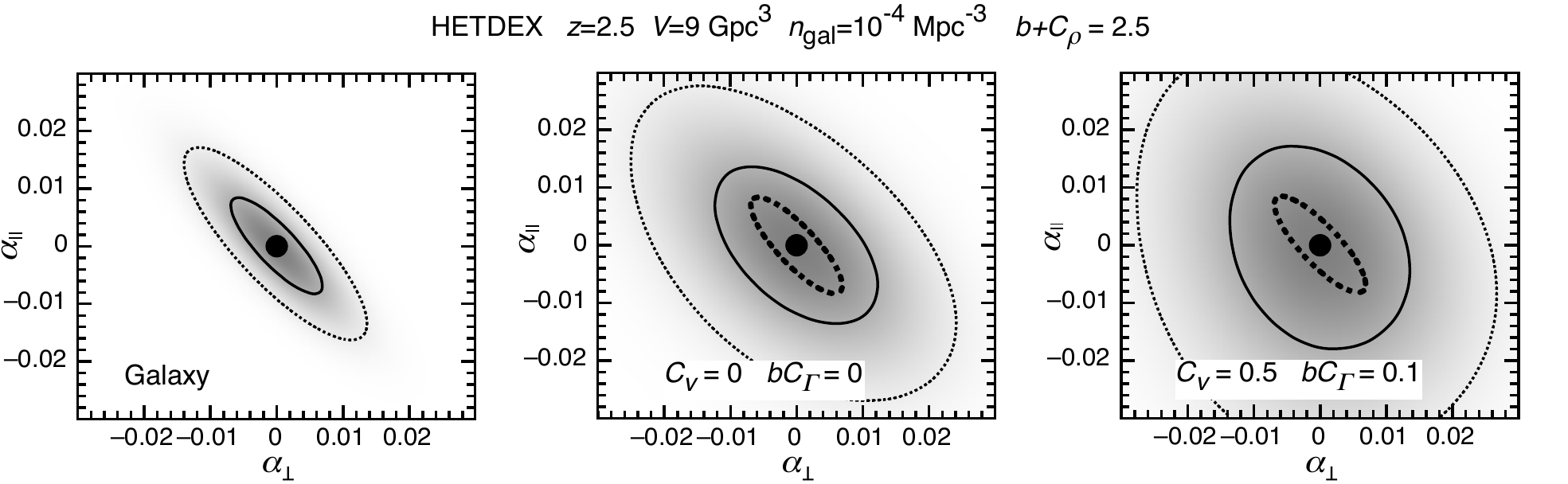}
\caption{Constraints on power-spectrum distortions achievable via the Alcock-Paczynski test. The three panels show contours of likelihood for the parameter set $(\alpha_\perp,\alpha_\parallel)$, in the cases of a traditional galaxy power-spectrum (left), and Ly$\alpha$ galaxy power-spectra with $[C_v,bC_\Gamma,(b+C_\rho)]=(0,0,2.5)$ (central) and $[C_v,bC_\Gamma,(b+C_\rho)]=(0.5,0.1,2.5)$ (right) respectively. Contours of likelihood for the values of $\alpha_\perp$ and $\alpha_\parallel$ are shown (at 61\% and 14\% of  the peak likelihood).  The 61\% contour for the traditional galaxy redshift survey constraints is repeated in the central and right panels for comparison (thick dotted line).}
\label{fig9}
\end{figure*}

\citet[][]{shoji2009} have argued that much more accurate constraints on cosmological distances are obtained by consideration of the whole power-spectrum shape rather than only the BAO scale. To quantify the potential of the Ly$\alpha$ galaxy power-spectrum for measuring cosmological parameters, we therefore calculate the Alcock-Paczynski effect~\citep[][]{Alcock1979}. Our approach is to specify the general result of \citet[][]{barkana2006} for the distortion of the true power-spectrum [$P_{\rm Ly}^{\rm t}(k,\mu)$] that results from an incorrect choice of cosmology. Dilation parameters $\alpha$ and $\alpha_\perp$ are used describe the distortions between the transverse and line-of-sight scales, and in the overall scale, respectively. These are defined such that $(1+\alpha)$ is the ratio between the assumed and true values of ($D_{\rm A}H$), while $(1+\alpha_\perp)$ is the ratio between the assumed and true values of the angular diameter distance, $D_{\rm A}$. In the Alcock-Paczynski test, the correct cosmology is inferred by finding cosmological parameters for which $\alpha=\alpha_\perp=0$.

To calculate the Alcock-Paczynski effect we apply equation~(8) in \citet[][]{barkana2006} 
\begin{eqnarray}
\label{AP}
\nonumber
P_{\rm Ly}(k,\mu)&=&(1+\alpha-3\alpha_\perp)P_{\rm Ly}^{\rm t}+(\alpha\mu^2-\alpha_\perp)\frac{\partial P_{\rm Ly}^{\rm t}}{\partial \ln{k}}\\
&&\hspace{26mm}+\alpha(1-\mu^2)\frac{\partial P_{\rm Ly}^{\rm t}}{\partial\ln{\mu}}
\end{eqnarray}
to the power-spectrum in equation~(\ref{PSmu}). Here the power-spectra $P_{\rm Ly}$ and
$P_{\rm Ly}^{\rm t}$ are evaluated at the observed $\vec{k}$. This
procedure results in a modified power-spectrum that is related to the true mass density power-spectrum
($P_{\rm m}^{\rm t}$) via
\begin{eqnarray}
\label{APPS}
\nonumber
P_{\rm Ly}(k,\mu) &=& P_{\rm m}^{\rm t} \left((b+C_\rho) + bC_\Gamma K(k)+(1-C_v)\mu^2\right)^2\\
\nonumber
&&\hspace{0mm}\times\left[(1+\alpha-3\alpha_\perp) + \frac{d\ln{P_{\rm m}^{\rm t}}}{d\ln{k}}(\alpha\mu^2-\alpha_\perp)\right]\\
\nonumber
&+& P_{\rm m}^{\rm t}  \left((b+C_\rho) + bC_\Gamma K(k)+(1-C_v)\mu^2\right)\\
\nonumber
&&\hspace{0mm}\times\left[2\frac{dK}{d\ln k} bC_\Gamma(\alpha\mu^2-\alpha_\perp)\right.\\ 
&&\hspace{15mm}\left.+4 \alpha(1-C_v)\left(1-\mu^2\right) f\mu^2 \right].
\end{eqnarray}

For dark energy studies it is more interesting to constrain the quantities $D_{\rm A}$ and $H$ independently rather than $D_{\rm A}$ and the product $D_{\rm A}H$. We define the dilation parameter $\alpha_\parallel$ such that $(1+\alpha_\parallel)$ is the ratio between the assumed values of $H$. The dilation parameter $\alpha$ in equation~(\ref{APPS}) is then expressed as 
\begin{equation}
\alpha = \frac{1+\alpha_\parallel}{1+\alpha_\perp}-1.
\end{equation}
Equation~(\ref{APPS}) can then be used to find the precision of constraints on $H$ and $D_{\rm A}$. Inspection of equation~(\ref{APPS}) suggests that some degeneracies are expected between cosmological constraints (parameterised by $\alpha_\parallel$ and $\alpha_\perp$) and the unknown parameters describing the modification to the power-spectrum owing to Ly$\alpha$ transmission fluctuations.

\subsection{Alcock-Paczynski constraints on the power-spectrum}

We next use equation~(\ref{APPS}) to calculate the permissible region of parameter space $\vec{p}=(\alpha_\parallel,\alpha_\perp,b+C_\rho,C_v,bC_\Gamma)$ around a true solution with power-spectrum $P^{\rm t}_{\rm Ly}$ and $\vec{p}_o= (0,0,2.5,0,0)$.  We have assumed that the mean-free-path of ionising photons is known a'priori, and do not fit it as a free parameter. We take the value to be $\lambda_{\rm mfp}=300$ co-moving Mpc \citep[][]{bolton2007,faucher-giguere2008}. Using the power-spectrum sensitivity specified in equation~(\ref{PSnoise}), we construct likelihoods
\begin{eqnarray}
\label{likelihood}
\nonumber
\ln{\mathcal{L}(\vec{p})} =& -&\frac{1}{2}\sum_{k,\mu}\left(\frac{P_{\rm Ly}(k,\mu,\vec{p})-P_{\rm Ly}^{\rm t}(k,\mu,\vec{p}_o)}{\Delta P_{\rm Ly}(k,\mu)}\right)^2\\
&+& \ln{\mathcal{L}_{C_\rho}} + \ln{\mathcal{L}_{C_v}} + \ln{\mathcal{L}_{C_\Gamma}}, 
\end{eqnarray}
where the sum is over bins of $k$ and $\mu$, and $\mathcal{L}_{C_\rho}$, $\mathcal{L}_{C_v}$ and  $\mathcal{L}_{C_\Gamma}$ are the a-priori likelihoods for the parameters $b+C_\rho$,  $C_v$ and $bC_\Gamma$ respectively. To account for the possibility of non-linearity in the smooth power-spectrum at small scales we restrict our fitting to wave numbers $k_{\rm max}<0.4$Mpc$^{-1}$.

\begin{figure*}
\includegraphics[width=17cm]{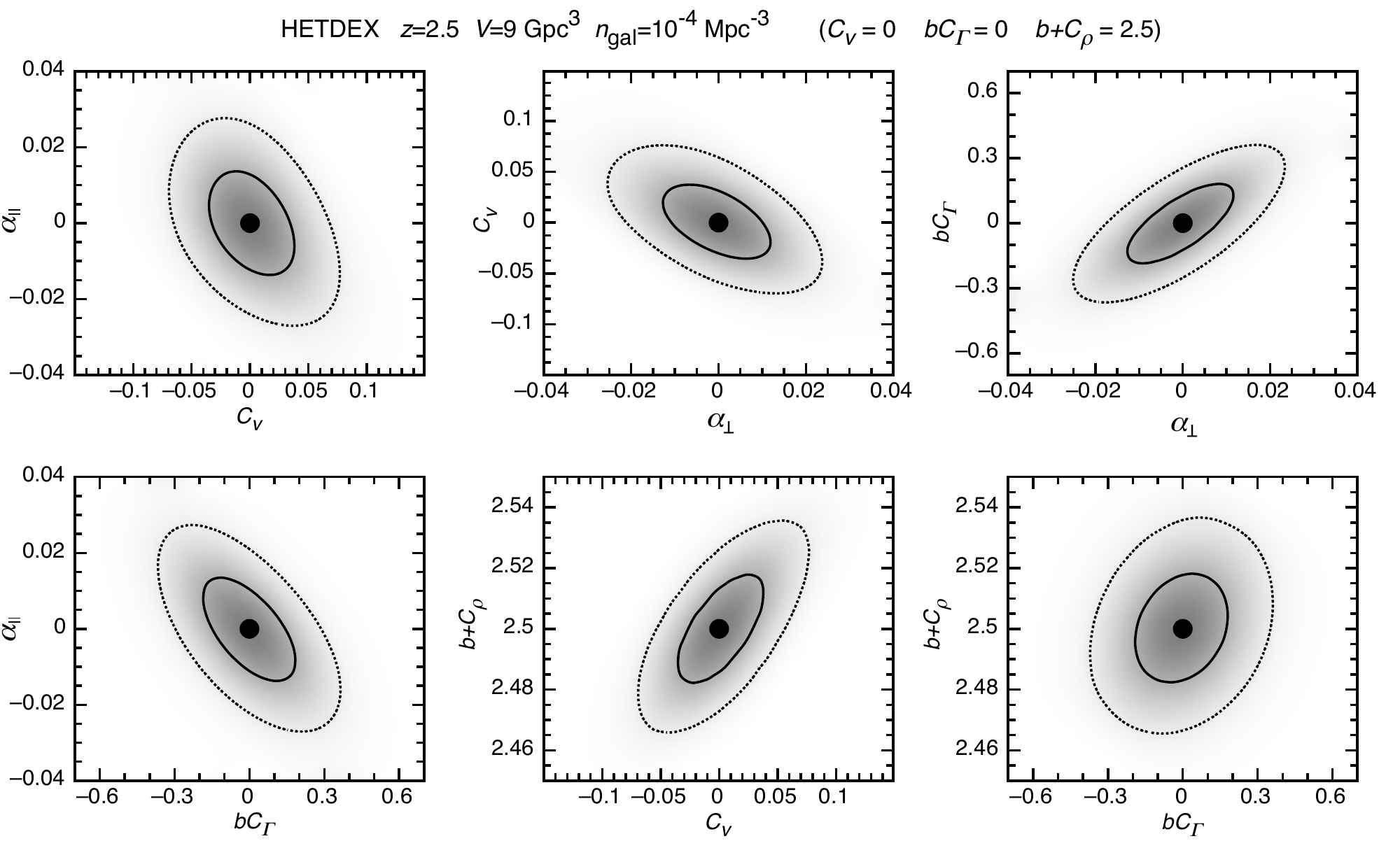}
\caption{Constraints on power-spectrum distortions and transmission models achievable via the Alcock-Paczynski test. The six panels show contours of likelihood for the parameter sets $(C_v,\alpha_\parallel)$, $(\alpha_\perp,C_v)$, $(\alpha_\perp,bC_\Gamma)$, $(bC_\Gamma,\alpha_\parallel)$, $(C_v,b+C_\rho)$ and $(bC_\Gamma,b+C_\rho)$, in the case of a Ly$\alpha$ galaxy power-spectrum with $[C_v,bC_\Gamma,(b+C_\rho)]=(0,0,2.5)$. Contours of likelihood for the values of $\alpha_\parallel$ and $\alpha_\perp$ are shown (at 61\% and 14\% of  the peak likelihood).}
\label{fig10}
\end{figure*}

\begin{figure*}
\includegraphics[width=17cm]{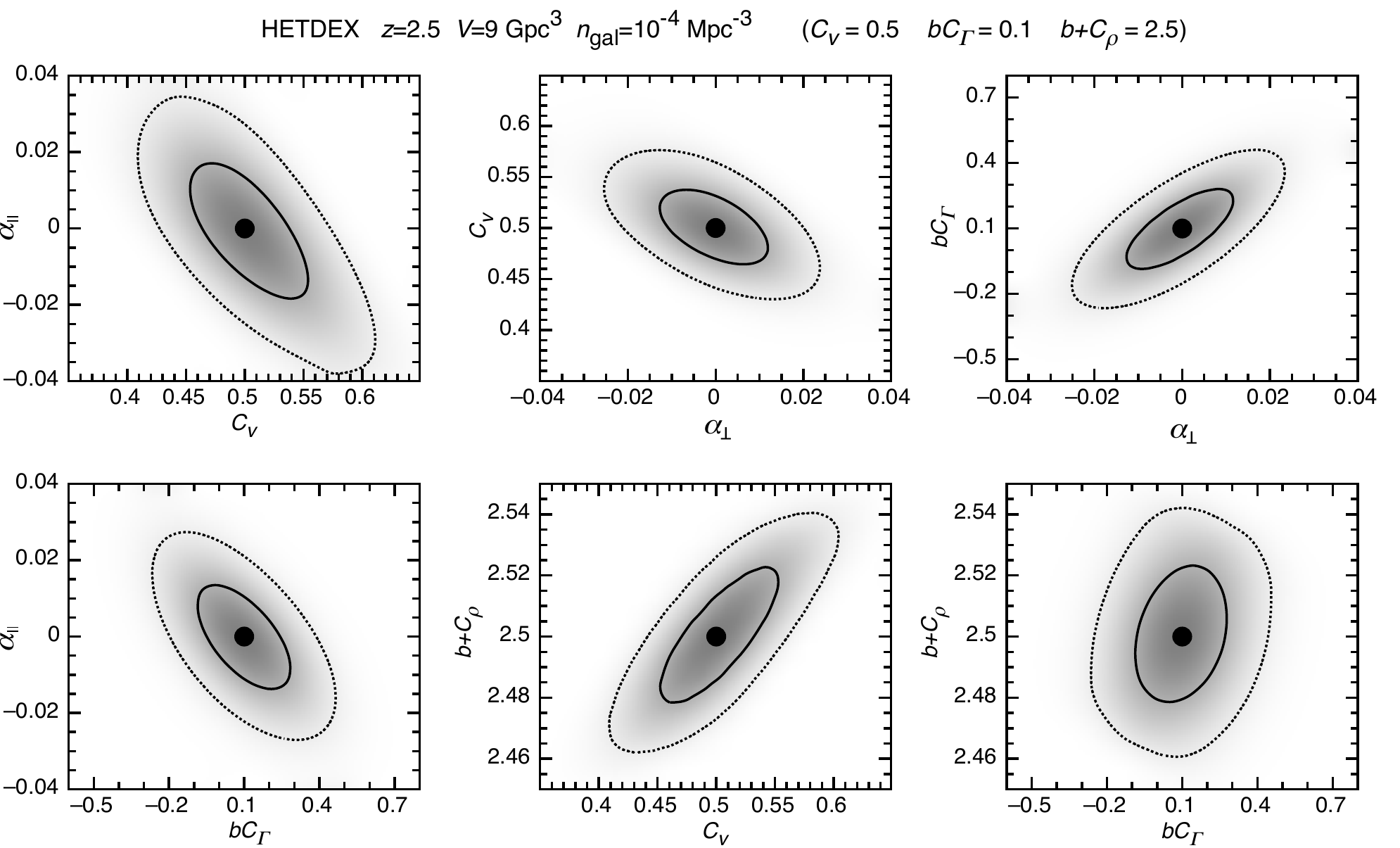}
\caption{Constraints on power-spectrum distortions and transmission models achievable via the Alcock-Paczynski test. As per Figure~\ref{fig10}, but for the case of a Ly$\alpha$ galaxy power-spectrum with $[C_v,bC_\Gamma,(b+C_\rho)]=(0.5,0.1,2.5)$.}
\label{fig11}
\end{figure*}

\subsection{Constraints for a traditional galaxy redshift survey}

To provide a baseline for our analysis we first consider the case where Ly$\alpha$ transmission has no effect on observed galaxy flux (i.e. we consider only power-spectra with $C_\rho=C_v=C_\Gamma=0$), as is the case for a traditional galaxy redshift survey. For this analysis we further assume that the value of $f$ is known and that the shape of the primordial power-spectrum is well measured by other means. We refer to these constraints as being for a {\em traditional galaxy redshift survey} in the remainder of this paper. As discussed in \S~\ref{shojicomp}, \citet[][]{shoji2009} have investigated the consequences of relaxing these strict prior constraints. 

As noted in \S~\ref{secAP}, the parameters $\alpha_\parallel$ and $\alpha_\perp$ are defined such that $(1+\alpha_\parallel)$ is the ratio between the assumed and true values of ($H$), while $(1+\alpha_\perp)$ is the ratio between the assumed and true values of the angular diameter distance, $D_{\rm A}$. Thus, the precision with which $\alpha_\perp$ can be measured provides an estimate of the relative precision with which $D_{\rm A}$ can be measured. Similarly, precision with which the local value of $H$ can be measured is provided by the precision with which $\alpha_\parallel$ can be measured. 

The left hand panel of Figure~\ref{fig9} presents likelihood contours for the parameter set $(\alpha_\perp,\alpha_\parallel)$ obtained from a traditional galaxy redshift survey assuming the HETDEX volume and galaxy density. The likelihood is marginalised over the bias $b$ assuming a flat prior probability. The contours show a degeneracy between $\alpha_\parallel$ and $\alpha_\perp$. This negative correlation is fundamental to the Alcock-Paczynski effect. When the redshift space distortion is known perfectly well, the departure of the real-space power spectrum from isotropy can be used to determine $D_{\rm A}H$ (or $\alpha$ in equation~\ref{AP}).  Figure~\ref{fig9} indicates that line-of-sight and angular distortions of the power-spectrum compared with an assumed model can be measured at better than the $\sim1\%$ level, indicating that our analysis is consistent with expectations for the HETDEX survey. Comparison with Figure~\ref{fig8} shows that the precision available on the line-of-sight and radial distances measured from the Alcock-Paczynski test using the full power-spectrum shape are a factor of several better than from an analysis of the BAO scale alone \citep[][]{shoji2009}. Values for these and subsequent constraints on $\alpha_\parallel$ and $\alpha_\perp$ are listed in Table~\ref{tab2}.

\begin{figure*}
\includegraphics[width=17cm]{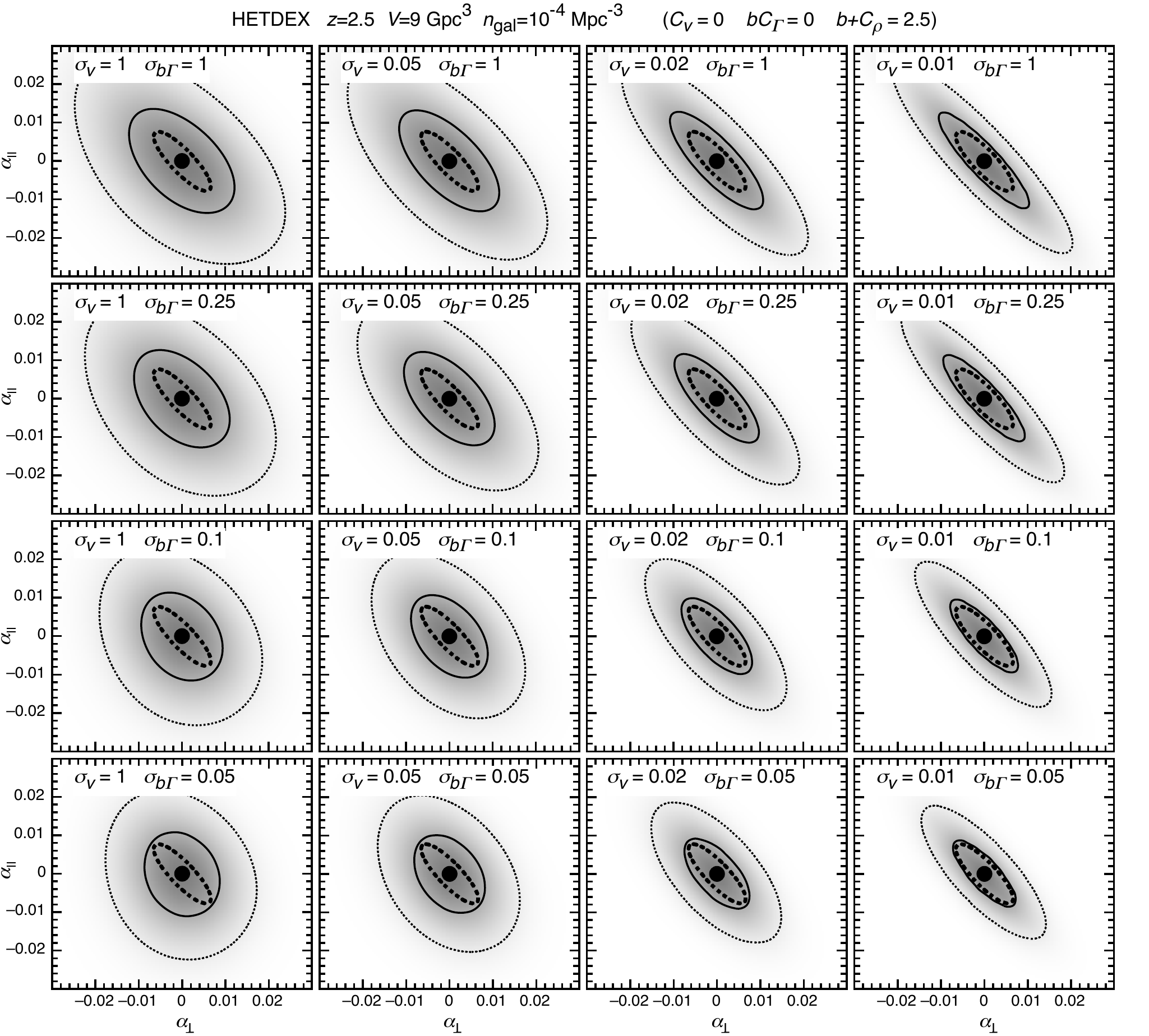}
\caption{Constraints on power-spectrum distortions achievable via the Alcock-Paczynski test including prior constraints on $C_v$ and $bC_\Gamma$. Each panel shows contours of likelihood for the parameter set $(\alpha_\perp,\alpha_\parallel)$, in the case of a Ly$\alpha$ galaxy power-spectrum with $[C_v,bC_\Gamma,(b+C_\rho)]=(0,0,2.5)$. Contours of likelihood for the values of $\alpha_\parallel$ and $\alpha_\perp$ are shown (at 61\% and 14\% of  the peak likelihood). Prior likelihoods of $\mathcal{L}_{C_v}(C_v)=\exp{(-(C_v-\langle C_v\rangle)^2/2\sigma_v^2)}$ and $\mathcal{L}_{C_\Gamma}(bC_\Gamma)=\exp{(-(bC_\Gamma-\langle bC_\Gamma\rangle)^2/2\sigma_{b\Gamma}^2)}$ were included in the constraints. The four columns show constraints assuming $\sigma_v=1$, 0.05, 0.02 and 0.01.  For each of these columns, four rows are shown with $\sigma_{b\Gamma}=1$, 0.25, 0.1 and 0.05. The 61\% contour for the traditional galaxy redshift survey constraints is plotted in each panel for comparison (thick dotted lines).}
\label{fig12}
\end{figure*}

\subsection{Constraints for a Ly$\alpha$ galaxy redshift survey}
\label{LyGalSur}

We next calculate the level to which Ly$\alpha$ transmission fluctuations influence the precision with which $\alpha_\parallel$ and $\alpha_\perp$ can be measured. The central panel of Figure~\ref{fig9} presents likelihood contours for the parameter set $(\alpha_\perp,\alpha_\parallel)$ obtained from a Ly$\alpha$ galaxy survey, again assuming the HETDEX volume and galaxy density, with a fiducial model having $[C_v,bC_\Gamma,(b+C_\rho)]=(0,0,2.5)$. Here the likelihood is marginalised over the parameters $C_v$, $bC_\Gamma$ and $(b+C_\rho)$ assuming flat prior probabilities (i.e. $\mathcal{L}_{C_\rho}=\mathcal{L}_{C_v}=\mathcal{L}_{C_\Gamma}=const$). The 1-sigma contour for the traditional galaxy redshift survey case is repeated for comparison. Figure~\ref{fig9} indicates that without knowledge of the detailed properties of Ly$\alpha$ transmission (enabling prediction of $C_v$, $C_\Gamma$ and $C_\rho$), the line-of-sight and angular distortions of the power-spectrum can be measured at the $\sim1.3\%$ level, a factor of $\sim1.5$ decrease in the available cosmological precision relative to a traditional galaxy redshift survey.

To calculate the constraints on $\alpha_\parallel$ and $\alpha_\perp$ for a Ly$\alpha$ galaxy survey we must specify the transmission model. In the above calculation we assumed a fiducial model with no transmission effects, but allowed for their possible existence when performing the cosmological fit. In the right panel of Figure ~\ref{fig9} we repeat this analysis for an assumed fiducial model which has $[C_v,bC_\Gamma,(b+C_\rho)]=(0.5, 0.1, 2.5)$, and so includes strong modification of the observed power-spectrum from fluctuations in Ly$\alpha$ transmission. In this case we find that constraints on the angular distortions of the power-spectrum are unchanged ($\sim1.3\%$), but that the line-of-sight distortions can only be measured at the $\sim1.7\%$ level.

To investigate the origin of the decrease in precision that follows inclusion of Ly$\alpha$ transmission fluctuations in a power-spectrum analysis, we show contours of likelihood for the parameter sets $(C_v,\alpha_\parallel)$, $(\alpha_\perp,C_v)$, $(\alpha_\perp,bC_\Gamma)$ and $(bC_\Gamma, \alpha_\parallel)$, in the cases of $[C_v,bC_\Gamma,(b+C_\rho)]=(0.5,0.1,2.5)$ and $[C_v,bC_\Gamma,(b+C_\rho)]=(0,0,2.5)$ (Figures~\ref{fig10} and \ref{fig11} respectively). For each set the likelihood is marginalised over the remaining parameters assuming flat prior probabilities. The figures show strong degeneracies as apparent from equation~(\ref{APPS}). These degeneracies represent the origin of the weakened constraints on $\alpha_\parallel$ and $\alpha_\perp$, which arise because the transmission fluctuations introduce scale and angular dependencies into the power-spectrum that mimic those introduced by an incorrect choice off cosmology. In particular, Figure~\ref{fig9} shows that the correlation between $\alpha_\parallel$ and $\alpha_\perp$ nearly disappears. This is because by marginalising over $C_v$, the redshift space distortions are now not perfectly known (as was assumed for the traditional galaxy redshift survey case), so that the Alcock-Paczynski test cannot be used to measure $D_{\rm A}H$.

We have also computed the contours of likelihood for the parameter sets $(C_v,b+C_\rho)$ and $(bC_\Gamma,b+C_\rho)$, in the cases of $[C_v,bC_\Gamma,(b+C_\rho)]=(0.5,0.1,2.5)$ and $[C_v,bC_\Gamma,(b+C_\rho)]=(0,0,2.5)$ (Figures~\ref{fig10} and \ref{fig11} respectively). These can be considered in addition to the constraints $(C_v,\alpha_\parallel)$, $(\alpha_\perp,C_v)$, $(\alpha_\perp,bC_\Gamma)$, $(bC_\Gamma,\alpha_\parallel)$,  and represent estimates for the available constraints on transmission models.  For each set the likelihood is marginalised over the remaining parameters assuming flat prior probabilities. The parameters $C_v$ and $bC_\Gamma$ describing the transmission model show very little degeneracy with each other. However  constraints on  $C_v$ are degenerate with the power-spectrum amplitude $(b+C_\rho)$. The constraints on these parameters are similar in magnitude for the two fiducial models considered.  Without detailed prior knowledge of the cosmology (i.e. no prior on $\alpha_\parallel$ or $\alpha_\perp$), a Ly$\alpha$ survey like HETDEX could determine uncertainties in $C_v$, $bC_\Gamma$ and $(b+C_\rho)$ of $\Delta C_v\sim\pm0.04$, $\Delta bC_\Gamma\sim\pm0.2$ and $\Delta (b+C_\rho)\sim\pm0.02$. We discuss constraints on transmission models in more detail in \S~\ref{LYAconstraints}.

\begin{table*}
\begin{center}
\caption{\label{tab2} Summary of constraints on $\alpha_\parallel$ and $\alpha_\perp$ for the different survey's and analysis methods. The label "galaxy" refers to a traditional galaxy redshift survey, whereas "Ly$\alpha$ gal." refers to an analysis that includes marginalisation over $bC_\Gamma$ and $C_v$. The labels "BAO" and "AP" refer to constraints based on the BAO scale using equation~(\ref{BAOPS}) and the power-spectrum shape via equation~(\ref{APPS}) respectively. Where relevant the values of the fiducial model  $C_v$ and $bC_\Gamma$, and the corresponding prior constraints $\sigma_v$ and $\sigma_{b\Gamma}$ are listed.}
\begin{tabular}{ccccccccc}
\hline
                   &                           &\multicolumn{2}{c}{fiducial model} &  \multicolumn{2}{c}{prior constraint}  & & \multicolumn{2}{c}{cosmological constraints}  \\
 survey      & constraint         &$C_v$ &  $bC_\Gamma$ &  $\sigma_v$ & $\sigma_{b\Gamma}$  & & $\Delta\alpha_\perp$   & $\Delta\alpha_\parallel$    \\\hline
   galaxy   &    BAO         &    ---    &   ---     &      ---  &  ---      &   &  2.0\%        &   3.0\%    \\
   Ly$\alpha$ gal.   & BAO         &   0       &    0      &    ---    &  ---   & & 2.0\%   &      3.0\%             \\
   Ly$\alpha$ gal.   &   BAO       &     0.5     &    0.1      &    ---            & ---    & & 2.10\%  &       3.25\%          \\\\
   galaxy   &     AP        &     ---      &   ---       &       ---         &  ---   &  & 0.70\%  &   0.85\%         \\
   Ly$\alpha$ gal.   &    AP           &     0     &    0      &      ---         &  ---   & & 1.25\%   &  1.35\%                \\
   Ly$\alpha$ gal.   &    AP           &      0.5    &      0.1    &     ---           & ---    &  & 1.35\%  &     1.80\%            \\\\
   Ly$\alpha$ gal.   &    AP           &      0    &      0    &        1.0          &   1.0  & & 1.20\%   &   1.35\%            \\   
   Ly$\alpha$ gal.   &    AP           &      0    &      0    &         0.05         &   0.25  &   &  1.05\% &  1.10\%              \\
   Ly$\alpha$ gal.   &    AP           &      0    &      0    &          0.02        &     0.1   &   &   0.80\%    & 1.00\%    \\
    Ly$\alpha$ gal.   &    AP           &        0       &    0      &     0.01             &  0.05   & &  0.75\% &  0.90\%                \\\hline

\end{tabular}
\end{center}
\end{table*}

\subsection{Cosmological constraints including prior probabilities for $C_v$ and $bC_\Gamma$. }

We have shown that fluctuations in Ly$\alpha$ transmission decrease the precision with which the angular diameter distance and Hubble parameter can be measured at $z\sim2.5$, relative to measurements from a traditional galaxy redshift survey. Assuming no prior knowledge of $C_v$ and $bC_\Gamma$, the decrease is found to be a factor of 1.5-2. On the other hand, if the parameters describing the Ly$\alpha$ transmission fluctuations are separately constrained, then the degeneracies seen in Figures~\ref{fig10} and \ref{fig11} imply that the parameters $\alpha_\parallel$ and $\alpha_\perp$ will be measured with increased precision.  In this section we investigate the degree of prior knowledge regarding transmission of Ly$\alpha$ flux (as parameterised by $C_v$ and $bC_\Gamma$) that is required in order to achieve measurements of cosmological parameters with a precision that would be available in a traditional galaxy redshift survey.

Figure~\ref{fig12} shows the constraints on power-spectrum distortions that are achievable via the Alcock-Paczynski test in cases where there are prior constraints on $C_v$ and $bC_\Gamma$. Each panel shows contours of likelihood for the parameter set $(\alpha_\perp,\alpha_\parallel)$, in the case of a Ly$\alpha$ galaxy power-spectrum with $[C_v,bC_\Gamma,(b+C_\rho)]=(0,0,2.5)$. To compute these constraints we have assumed prior likelihoods of 
\begin{eqnarray}
\nonumber
\mathcal{L}_{C_v}(C_v)&=&\exp{(-(C_v-\langle{C}_v\rangle)^2/2\sigma_v^2)}\hspace{5mm}\mbox{and}\\
\mathcal{L}_{C_\Gamma}(bC_\Gamma)&=&\exp{(-(bC_\Gamma-\langle{bC}_\Gamma\rangle)^2/2\sigma_{b\Gamma}^2)},
\end{eqnarray}
where $\langle C_v\rangle$ and $\langle bC_\Gamma\rangle$ are the means of $C_v$ and $bC_\gamma$ corresponding to the fiducial model, and $\sigma_v$ and $\sigma_{b\Gamma}$ are the prior uncertainties in $C_v$ and $bC_\Gamma$ respectively.
Figure~\ref{fig12} is presented as a grid, with prior precision on $C_v$ increasing from left to right, and prior precision on $bC_\Gamma$ increasing from top to bottom.  The four columns show constraints assuming $\sigma_v=1$, 0.05, 0.02 and 0.01.  For each of these columns, four rows are shown with $\sigma_{b\Gamma}=1$, 0.25, 0.1 and 0.05. The 61\% contour for the traditional galaxy redshift survey constraints is repeated in all panels for comparison (thick dotted lines). 

As the prior precision on $bC_\Gamma$ is increased, the precision with which $\alpha_\perp$ is measured increases. As the prior precision on $C_v$ is increased, the strength of the correlation between $\alpha_\parallel$ and $\alpha_\perp$ is increased. This signifies that the parameter $C_v$ is degenerate with the Alcock-Paczynski effect. As a result, precision in $\alpha_\parallel$ requires prior knowledge of both $bC_\Gamma$ and $C_v$. Prior uncertainties with values smaller than $\sigma_{b\Gamma}\la0.05$ and $\sigma_v\la0.01$ provide sufficient precision that fluctuations in transmission do not dominate the uncertainties in the clustering, in which case a Ly$\alpha$ galaxy survey could be used to measure cosmological parameters with a precision close to that available in a traditional galaxy redshift survey.

\subsection{Comparison with previous HETDEX forecasts}
\label{shojicomp}

\citet[][]{shoji2009} have presented forecasts for a galaxy redshift survey with parameters corresponding to HETDEX. Comparison with the results of their study serves both as a check of our analysis, and illustrates the relationship between the astrophysical parameters introduced through $C_\rho$, $C_\Gamma$ and $C_v$, and the cosmological parameters $\alpha_\parallel$ and $\alpha_\perp$. The results from the analysis of \citet[][]{shoji2009} are listed in their Table~1, and assume a value of bias $b=2.5$, making the constraints directly comparable to this paper. Firstly, our case of a traditional galaxy power-spectrum should be compared to the constraints on $\alpha_\parallel$ and $\alpha_\perp$ obtained where constraints are marginalised only over the power-spectrum amplitude. \citet[][]{shoji2009} find $\Delta \alpha_\parallel=0.78\%$ and 
$\Delta \alpha_\perp=0.88\%$ in this case, which corresponds well to our values of $0.85\%$ and $0.7\%$.

In order to compare our results for a Ly$\alpha$ selected galaxy redshift survey to the work of \citet[][]{shoji2009}, we first consider the case of the limit where $k\lambda\ll1$, for which our power spectrum model is 
\begin{equation}
\nonumber
P_{\rm Ly\alpha}(k,\mu)=P(k)\left[ (b+C_{\rho}+bC_\Gamma) + (1-C_v)f\mu^2 \right]^2.
\end{equation}
Defining the power-spectrum with a new bias parameter such that $B\equiv (b+C_{\rho}+bC_\Gamma)$ and the \citet[][]{kaiser1987} factor such that $F\equiv(1-C_v)f$, we obtain
\begin{equation}
P_{\rm Ly\alpha}(k,\mu)=P(k)\left[B+F\mu^2\right]^2,
\end{equation}
which is identical to the usual form. The modifications to the bias and Kaiser factor should therefore not affect the Alcock-Pacynski effect. However the addition of transmission fluctuations means that constraints must be marginalised over the Kaiser factor, which changes the shape of contours. 

This can be seen in Figure 3 of \citet[][]{shoji2009}, where examples of constraints are shown that include the cases of  marginalisation over amplitude, and also of marginalisation over both amplitude and the \citet[][]{kaiser1987} factor. Marginalising over the \citet[][]{kaiser1987} factor introduces additional uncertainty, with constraints of $\Delta \alpha_\parallel=1.13\%$ and $\Delta \alpha_\perp=1.10\%$ in this case. These values should be compared to our constraints  of $\Delta \alpha_\parallel=1.1\%$ and $\Delta \alpha_\perp=0.9\%$ which are obtained with a broad prior on $C_v$, but tight constraints on $C_\gamma$ (see Figure~\ref{fig12}). Finally, including the scale dependent ionizing back-ground term $K(k)$ represents a similar effect to that of the primordial spectral index and its running. \citet[][]{shoji2009} presented constraints that included marginalising over the amplitude, \citet[][]{kaiser1987} factor, and primordial power-spectrum shape, finding $\Delta \alpha_\parallel=1.36\%$ and $\Delta \alpha_\perp=1.23\%$. These values are again very similar to our constraints without priors on $C_\rho$, $C_\Gamma$ or $C_v$ of $\Delta \alpha_\parallel=1.35\%$ and  $\Delta \alpha_\perp=1.25\%$

Thus, the level at which transmission fluctuations will influence measurements of cosmological distance are comparable to those from marginalising over other cosmological parameters.   However interpretations of measured quantities like the power-spectrum shape or the growth function $f$ will need to account for the astrophysical effects of Ly$\alpha$ transmission. For example, the possibility of a non-zero $C_v$ will complicate interpretation of the extracted value of the redshift distortion factor, because the measurement will be of $(1-C_v)f$, rather than of $f$. This degeneracy will make it very difficult to test theories of modified gravity theory using the redshift space distortion~\citep[e.g.][]{blake2010}.

\section{Constraints on Ly$\alpha$ transmission models}

\begin{figure*}
\includegraphics[width=17cm]{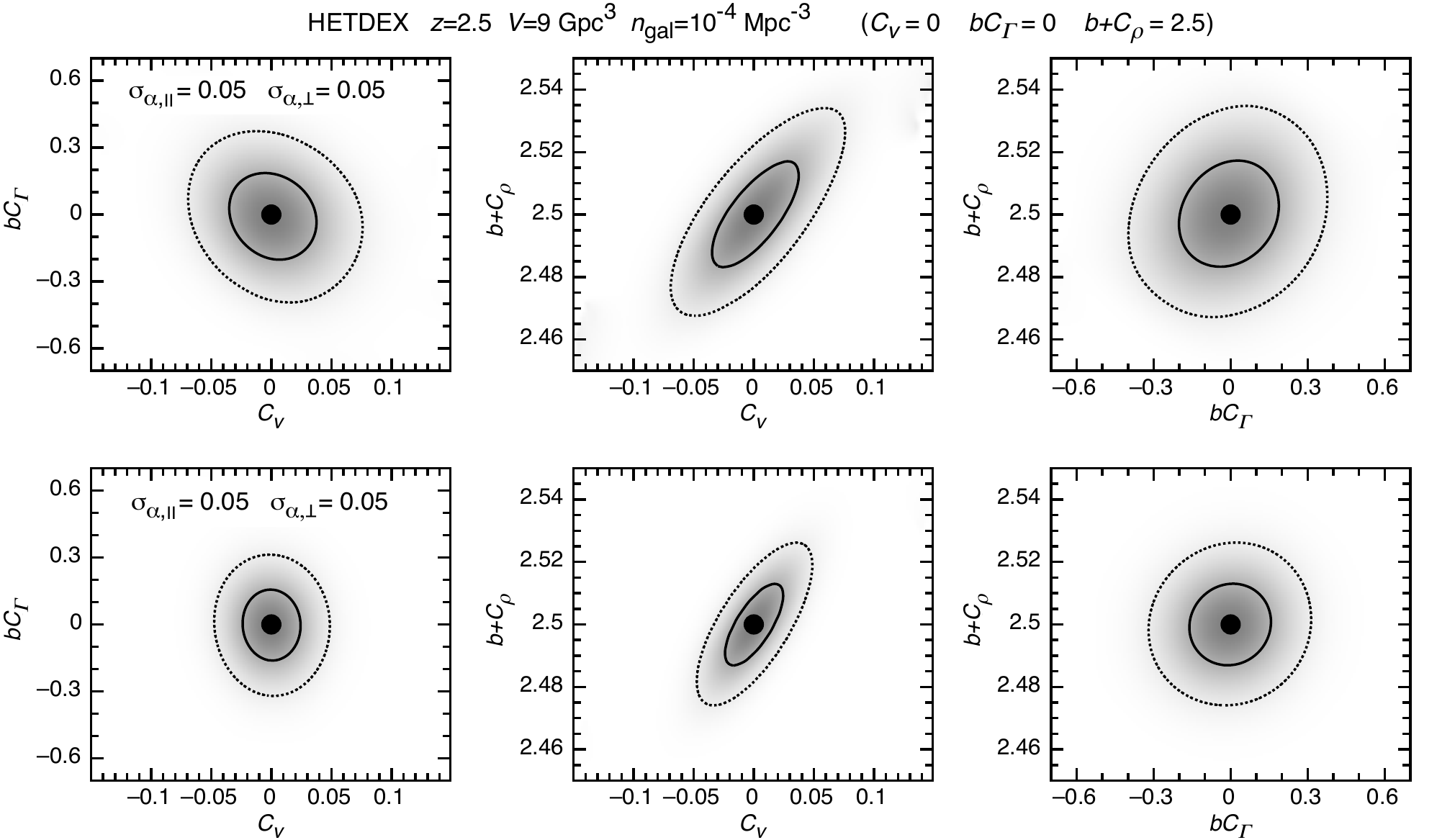}
\caption{Constraints on the parameters $(b+C_\rho)$, $C_\Gamma$ and $C_v$ based on power-spectrum distortions. Uncertainties on the cosmology are included via the Alcock-Paczynski test. The left, central and right panels show contours of likelihood for the parameter sets $(C_v,bC_\Gamma)$, $(C_v,b+C_\rho)$ and $(bC_\Gamma,b+C_\rho)$, in the case of a Ly$\alpha$ galaxy power-spectrum with $[C_v,bC_\Gamma,(b+C_\rho)]=(0,0,2.5)$. Contours of likelihood for the values of $\alpha_\parallel$ and $\alpha_\perp$ are shown (at 61\% and 14\% of  the peak likelihood). We have included prior likelihoods of $\mathcal{L}_{\alpha}(\alpha_\parallel)=\exp{(-\alpha_\parallel^2/2\Delta \alpha_\parallel^2)}$ and $\mathcal{L}_{\alpha_\perp}(\alpha_\perp)=\exp{(-\alpha_\perp^2/2\Delta\alpha_\perp^2)}$ to represent different precisions of knowledge of the cosmology at z=2.5. In the upper and lower panels $\Delta\alpha=\Delta\alpha=0.05$ and $\Delta\alpha=\Delta\alpha=0.01$ respectively.}
\label{fig13}
\end{figure*}
\label{LYAconstraints}

We have shown that fluctuations in Ly$\alpha$ transmission will result in modification of the scale and angular dependance of the power-spectrum measured from a large Ly$\alpha$ galaxy redshift survey like HETDEX. These modifications in the power-spectrum potentially reduce the ability of the measured power-spectrum to constrain cosmological parameters. On the other hand, our study also shows that the parameters describing the modifications to power-spectrum are measured as part of the fitting process. Before concluding, we therefore calculate the available precision on the parameters $b+C_\rho$, $bC_{\rm \Gamma}$ and $C_v$, whose values provide insight into the level of suppression of Ly$\alpha$ flux, and the astrophysics of the interaction between Ly$\alpha$ emitting galaxies and the IGM (\S~\ref{detailedmodel}).

In Figure~\ref{fig13} we present constraints on the parameter sets $(C_v,bC_\Gamma)$, $(C_v,b+C_\rho)$ and $(bC_\Gamma,b+C_\rho)$, in the case of a Ly$\alpha$ galaxy power-spectrum with $[C_v,bC_\Gamma,(b+C_\rho)]=(0,0,2.5)$. The numerical values of the measured uncertainties are listed in Table~\ref{tab3}. These constraints assume that the cosmological model is measured from other sources. The uncertainties in the cosmology are included in our analysis via the Alcock-Paczynski test (equation~\ref{APPS}), which for this application provides a measure of the uncertainty in the power-spectrum shape through the dilation parameters $\alpha_\parallel$ and $\alpha_\perp$. An exception is that the uncertainty in the mass power-spectrum amplitude (proportional to the normalisation of the primordial power-spectrum, $\sigma_8$) which is degenerate with $(b+C_\rho)$. Our analysis assumes that the shape of the primordial power-spectrum and the value of $f$ are known. 

Using the power-spectrum sensitivity specified in equation~(\ref{PSnoise}), we construct likelihoods
\begin{eqnarray}
\label{likelihood}
\nonumber
\ln{\mathcal{L}(\vec{p})} =& -&\frac{1}{2}\sum_{k,\mu}\left(\frac{P_{\rm Ly}(k,\mu,\vec{p})-P_{\rm Ly}^{\rm t}(k,\mu,\vec{p}_o)}{\Delta P_{\rm Ly}(k,\mu)}\right)^2\\
\nonumber
&+& \ln{\mathcal{L}_{C_\rho}} + \ln{\mathcal{L}_{C_v}} + \ln{\mathcal{L}_{C_\Gamma}}\\
&+& \ln{\mathcal{L}_{\alpha_\parallel}} + \ln{\mathcal{L}_{\alpha_\perp}} ,
\end{eqnarray}
where the sum is over bins of $k$ and $\mu$. We assume flat prior probabilities for the transmission dependent parameters (i.e. $\mathcal{L}_{C_\rho}=\mathcal{L}_{C_v}=\mathcal{L}_{C_\Gamma}=const$). To quantify the uncertainty we have included prior likelihoods of $\mathcal{L}_{\alpha_\parallel}(\alpha_\parallel)=\exp{(-\alpha_\parallel^2/2\sigma_{\alpha,\parallel}^2)}$ and $\mathcal{L}_{\alpha_\perp}(\alpha_\perp)=\exp{(-\alpha_\perp^2/2\sigma_{\alpha,\perp}^2)}$. In the upper and lower panels of Figure~\ref{fig13} we assume uncertainties of $\sigma_{\alpha,\parallel}=\sigma_{\alpha,\perp}=0.05$ and $\sigma_{\alpha,\parallel}=\sigma_{\alpha,\perp}=0.01$ respectively.      

Given cosmological uncertainties $\sigma_{\alpha,\parallel}=\sigma_{\alpha,\perp}=0.05$, we find that the values of $b+C_\rho$, $bC_\Gamma$ and $C_v$ could be constrained with  precisions of $\Delta(b+C_\rho)\sim\pm0.02$, $\Delta (bC_\Gamma)\sim\pm0.25$ and $\Delta C_v\sim\pm0.04$. For next generation cosmological constraints with  $\Delta\alpha_\parallel=\Delta\alpha_\perp=0.01$ we find smaller errors on Ly$\alpha$ clustering parameters of  $\Delta(b+C_\rho)\sim\pm0.015$, $\Delta (bC_\Gamma)\sim\pm0.15$ and $\Delta C_v\sim\pm0.02$. These errors should be compared to the predicted values in Table~\ref{tab1}, and with the results of Figure~\ref{fig4}. Since the measurement of non-zero values of the parameters $b+C_\rho$, $bC_\Gamma$ or $C_v$ indicates that the Ly$\alpha$ line is partially absorbed in the IGM (i.e. $F>0$ and $\tau>0$ in the analytic model), the available constraints in a survey like HETDEX could easily measure the presence of Ly$\alpha$ absorption in the IGM. 

\begin{table*}
\begin{center}
\caption{\label{tab3} Summary of constraints on $(b+C_\rho)$ $bC_\Gamma$ and $C_v$. The values of the fiducial model $(b+C_\rho)$, $bC_\Gamma$ and $C_v$, and the prior constraints  $\alpha_\parallel$ and $\alpha_\perp$  are listed.}
\begin{tabular}{ccccccccc}
\hline
 \multicolumn{2}{c}{fiducial model} &  \multicolumn{2}{c}{prior constraint}  & & \multicolumn{3}{c}{Ly$\alpha$ transmission constraints}  \\
 $bC_\Gamma$ & $C_v$ &  $\sigma_{\alpha,\perp}$   & $\sigma_{\alpha,\parallel}$  & &  $\Delta(b+C_\rho)$ & $\Delta(bC_\Gamma)$ & $\Delta(C_v)$ &   \\\hline
    0    &   0    &      0.05 & 0.05    &   &  0.017        &   0.19   &  0.040  \\
       0       &    0      &     0.01             &  0.01   & &  0.013\ &  0.15    &  0.025           \\\hline

\end{tabular}
\end{center}
\end{table*}

Of the available constraints, the parameter $(b+C_\rho)$ cannot be used to constrain transmission models despite the high precision with which it will be determined. This is because of the degeneracy between $C_\rho$ and the unknown galaxy bias $b$.  Our models predict small values of $bC_\Gamma$ (see \S~\ref{coefficients} and \S~\ref{detailedmodel}), except in very special cases, and so the available precision will not provide useful constraints. However the precision $\Delta C_v\sim\pm0.04$ is small compared with expected values of $C_v$ for a range of scenarios \citep[\S~\ref{detailedmodel}, see also][]{zheng2010}.  In particular, the HETDEX survey could distinguish between infall and outflow dominated models of IGM transmission, for which the parameters range from 0.05 to 0.7. Measurement of this term would directly determine the extent to which the IGM impacts the observed flux, providing critical  information on the intrinsic Ly$\alpha$ luminosity, and the presence of outflows.

\section{Summary and Conclusion}
\label{conclusions}

Wide-field searches are now finding Ly$\alpha$ emitters in large numbers, and these galaxies contribute greatly to our understanding of the star-formation history, and of galaxy formation. In the near future, very large surveys of Ly$\alpha$ emitting galaxies will provide precision measurements of large scale clustering at $z\sim2.5$, and allow measurement of cosmological parameters at this previously unexplored epoch. However, to realise this goal it is crucial to understand the contribution to the observed clustering amplitude from fluctuations in inter-galactic absorption of intrinsic Ly$\alpha$ flux. In this paper we have shown that the environmental dependence of Ly$\alpha$ absorption can lead to significant non-gravitational features in the redshift space power-spectrum of Ly$\alpha$ galaxies, under a range of different physical scenarios. We have derived a physically motivated fitting formula that relates the scale and direction dependent Ly$\alpha$ power-spectrum to the mass power-spectrum [$P(k)$] 
\begin{eqnarray}
\nonumber
&&\hspace{-7mm}P_{\rm Ly\alpha}(k,\mu)=P(k)\\
&&\hspace{-2mm}\times\left( b\left(1+C_{\Gamma}\frac{\arctan{(k\lambda)}}{k\lambda}\right)+C_{\rho} + (1-C_v)f\mu^2 \right)^2.
\end{eqnarray}
This formula can be used in the power-spectrum  analyses of  a galaxy redshift survey to account for the environmental dependence of Ly$\alpha$ absorption, which includes fluctuations in density, ionising background and velocity gradient (parameterised by $C_\rho$, $C_\Gamma$ and $C_v$ respectively). We have presented a simple analytic model to calculate the values of these parameters. Our calculations imply that standard Ly$\alpha$ absorption scenarios will yield values for $C_\rho$, $C_\Gamma$ and $C_v$ that are at the tens of percent level, indicating that fluctuations in absorption of the Ly$\alpha$ line in the IGM can lead to modifications of the power-spectrum of Ly$\alpha$ selected galaxies that are of order unity. 

While our analytic model is useful for investigating the qualitative dependencies of clustering in Ly$\alpha$ selected galaxies, more detailed analyses are required to quantitatively predict the values of the constants $C_\rho$, $C_\Gamma$ and $C_v$. To quantify the expected effect of fluctuations in Ly$\alpha$ absorption on the observed power-spectrum we have therefore employed previously published models of Ly$\alpha$ radiative transfer. These models explore the combined effects of local star-formation and IGM infall, and galactic wind driven outflows on the transmission of the Ly$\alpha$ line through the circum-galactic IGM. We find that an infall dominated model for Ly$\alpha$ transmission predicts significant contributions (of order unity) to the observed power-spectrum. On the other hand, an outflow dominated model for Ly$\alpha$ transmission predicts contributions to the observed power-spectrum that are an order of magnitude smaller, at the level of $\sim5-10\%$. 

We have shown that the expected non-zero values of $C_\rho$, $C_\Gamma$ and $C_v$ will complicate attempts to use the clustering of Ly$\alpha$ emitters to constrain cosmological parameters in very large scale surveys. To quantify the influence of Ly$\alpha$ absorption on the ability of a large Ly$\alpha$ galaxy survey to constrain cosmological parameters, we have applied our modified redshift space power-spectrum to a survey with parameters corresponding to the planned HETDEX. We considered both cosmological constraints obtained from the BAO scale, and from the the full shape of the power-spectrum as measured by the Alcock-Paczynski effect. To base-line our study we also consider the case of a traditional galaxy redshift survey where the probability of galaxy selection is not a function of environment. Our analysis shows that  a survey with the parameters of HETDEX could measure the BAO scale along, and transverse to the line-of-sight with precisions of $\sim3\%$, and $\sim2\%$ respectively. We find that this precision is unaffected by modifications to the power-spectrum that arise from fluctuations in Ly$\alpha$ absorption. This finding is consistent with previous studies which have shown that sources of scale dependent bias can be removed in order to correctly recover the BAO scale. 

For a traditional galaxy redshift survey with the volume and galaxy density of HETDEX, much tighter constraints are available through consideration of the full power-spectrum shape \citep[][]{shoji2009}. In this case we find that find that the Alcock-Paczynski effect could be used to constrain both the line-of-sight and transverse directions at better than the 1\% level for a traditional galaxy redshift survey in which the shape of the primordial power-spectrum and the growth function $f$ are known. However, our analysis shows that the dependence of observed Ly$\alpha$ flux on velocity gradient and ionising background has the potential to compromise the cosmological information available from the full power-spectrum shape measured in a Ly$\alpha$ selected galaxy redshift survey. In a scenario where there is no prior knowledge of the details of Ly$\alpha$ absorption in the IGM, we find the precision of line-of-sight and transverse distance measurements the HETDEX would be decreased by a factor of 1.5-2,  from $\sim$0.8\% in the case of a traditional galaxy redshift survey, to $\sim1.3\%-1.7\%$. The weakened constraints on $\alpha$ and $\alpha_\perp$ arise because Ly$\alpha$ transmission fluctuations introduce scale and angular dependencies into the power-spectrum that are degenerate with those of an incorrect cosmology. In particular, the effect of a fluctuating ionizing background on the shape of the observed power-spectrum of Ly$\alpha$ selected galaxies is similar to that of an uncertainty in the shape of the primordial power-spectrum, while fluctuations in velocity gradient have an effect that is similar to that of redshift space distortions.

We also investigated the precision with which modelling of the Ly$\alpha$ radiative transfer must be understood in order for HETDEX to achieve a goal of better than 1\% distance measurements based on Ly$\alpha$ galaxy clustering. We find that as the prior precision on $C_\Gamma$ is increased, the accuracy with which $\alpha_\perp$ is measured also increases. As the prior precision on $C_v$ is increased, the strength of the correlation between $\alpha_\parallel$ and $\alpha_\perp$ is increased. However increased precision in $\alpha_\parallel$ requires prior knowledge of both $bC_\Gamma$ and $C_v$.  Prior uncertainties with values smaller than $\Delta C_\Gamma\la0.05$ and $\Delta C_v\la0.01$ provide sufficient accuracy that fluctuations in transmission do not dominate the uncertainties in the clustering.  In such cases a Ly$\alpha$ galaxy survey could be used to measure cosmological parameters with a precision comparable to a traditional galaxy redshift survey of equivalent volume and number density. On the other hand, these uncertainties are at a level below the accuracy with which $C_\rho$, $C_\Gamma$ and $C_v$ can be reliably predicted based on current theoretical understanding. 

We have turned the above analysis around, and assumed instead that the cosmology is known a-priori from other sources. The clustering of Ly$\alpha$ emitters can then be used to measure the impact of the IGM on observed Ly$\alpha$ lines, and to infer  the properties of the Ly$\alpha$ transmission model. Our models predict small values of $C_\Gamma$, and so the precision available from a survey like HETDEX ($\Delta C_\Gamma\sim\pm0.15-0.25$) will not provide useful constraints. However the precision  $\Delta C_v\sim\pm0.02-0.04$ would be small compared with the expected values for a range of scenarios (\S~\ref{detailedmodel}). For example, clustering of Ly$\alpha$ galaxies could distinguish between infall and outflow dominated models of IGM transmission. Measurement of this term in a survey like HETDEX would therefore directly determine the extent to which the IGM impacts the observed flux, providing information on the kinematics of the cold gas through which the Ly$\alpha$ photons are scattering.

The nature of Ly$\alpha$ emitters, their role in galaxy formation, and their utility as probes of cosmology, and of the state of the intergalactic and interstellar media are important topics to which large ongoing observational programs promise to make significant contributions. In this paper we have shown that power-spectrum measurements from a very large survey of Ly$\alpha$ selected galaxies could be used to study the relationship between the observed Ly$\alpha$ flux, and the astrophysics of the galaxy-IGM connection. However we also show that Ly$\alpha$ transmission fluctuations decrease the cosmological precision of a galaxy redshift survey by a factor of 1.5-2 relative to the best case precision available in a traditional galaxy redshift survey. Realising the full cosmological cosmological potential of a survey like HETDEX will therefore require a much more detailed theoretical understanding of the astrophysics that determines the relationship between the observed and intrinsic Ly$\alpha$ flux. Our study underlines the need for continued work to understand the role of radiative transfer through the inter-stellar and inter-galactic media in determining the observed properties of Ly$\alpha$ emission.

{\bf Acknowledgments} We thank Eiichiro Komatsu for very helpful insights, suggestions, and discussion of our results. JSWB thanks the Max Planck Institute fur Astrophysik for their hospitality during this work. The research was supported by the Australian Research Council (JSBW).

\bibliography{text}

\newcommand{\noopsort}[1]{}

\label{lastpage}
\end{document}